\documentclass[twocolumn,pra,showpacs,superscriptaddress,amssymb,amsmath,amsmath]{revtex4-1}
\usepackage{graphicx}
\usepackage{epstopdf}
\usepackage{comment}
\usepackage{hyperref}
\usepackage{bm}

\hypersetup{pdfstartpage=1,pdfmenubar=true,pdfpagemode=None,pdftoolbar=true,
urlcolor=blue,linkcolor=blue,colorlinks = true,citecolor=blue, bookmarksopen=false}

\begin{document}

\title{Interactions and chemical reactions in ionic alkali-metal and alkaline-earth-metal diatomic $AB^+$\\ and triatomic $A_2B^+$ systems}

\author{Micha\l~\'Smia{\l}kowski} 
\affiliation{Faculty of Physics, University of Warsaw, Pasteura 5, 02-093 Warsaw, Poland}
\affiliation{Faculty of Chemistry, University of Warsaw, Pasteura 1, 02-093 Warsaw, Poland}
\author{Micha\l~Tomza}
\email{michal.tomza@fuw.edu.pl}
\affiliation{Faculty of Physics, University of Warsaw, Pasteura 5, 02-093 Warsaw, Poland}

\date{\today}

\begin{abstract}

We theoretically characterize interactions, energetics, and chemical reaction paths in ionic two-body and three-body systems of alkali-metal and alkaline-earth-metal atoms in the context of modern experiments with cold hybrid ion-atom mixtures. Using \textit{ab initio} techniques of quantum chemistry such as the coupled-cluster method, we calculate ground-state electronic properties of all diatomic $AB^+$ and most of triatomic $A_2B^+$ molecular ions  consisting of Li, Na, K, Rb, Cs, Mg, Ca, Sr, Ba, and Yb atoms. Different geometries and wave-function symmetries of the ground state are found for different classes of molecular ions. We analyze intermolecular interactions in the investigated systems including additive two-body and nonadditive three-body ones. As an example we provide two-dimensional interaction potential energy surfaces for KRb$^+$+K and Rb$^+$+Sr$_2$ mixtures. We identify possible channels of chemical reactions based on the energetics of the reactants. The present results may be useful for investigating controlled chemical reactions and other applications of molecular ions formed in cold hybrid ion-atom systems.  

\end{abstract}

\maketitle

\section{Introducion}

The field of ultracold quantum matter has seen groundbreaking developments in recent years, unfolding perspectives for numerous applications in precision measurements~\cite{DeMilleScience17}, many-body physics~\cite{GrossScience17}, and controlled chemistry~\cite{BohnScience17} experiments. Laser cooling and trapping techniques have allowed one to prepare and manipulate ultracold atoms, ions, and molecules with full control over their quantum states, recently, also at the single particle level~\cite{WinelandRMP13,HarocheRMP13}. Further developments in experimental methods have opened the way for combining ultracold trapped ions and atoms in a single experimental setup~\cite{HarterCP14,CoteAAMOP16,TomzaRMP19}. Most of ion-atom experiments use alkaline-earth-metal ions trapped and laser-cooled in a Paul trap immersed into ultracold neutral alkali-metal or alkaline-earth-metal atoms trapped in magnetic, magneto-optical, or dipole traps~\cite{TomzaRMP19}. Alkaline-earth-metal ions and alkali-metal atoms are employed because of their electronic sturcture favorable for laser cooling. Several cold atomic ion-atom combinations have already been experimentally investigated, including: Yb$^+$+Yb~\cite{GrierPRL09}, Ca$^+$+Rb~\cite{HallMP13a}, Ba$^+$+Ca~\cite{SullivanPRL12}, Yb$^+$+Ca~\cite{RellergertPRL11}, Yb$^+$+Rb~\cite{ZipkesNature10}, Ca$^+$+Li~\cite{HazePRA13}, Ca$^+$+Rb~\cite{HallPRL11}, Ca$^+$+Na~\cite{SmithAPB14}, Yb$^+$+Li~\cite{JogerPRA17,FurstPRA18}, Sr$^+$+Rb~\cite{MeirPRL16}, Rb$^+$+Rb~\cite{RaviNatCommun12,HartePRL12}, Na$^+$+Na~\cite{SivarajahPRA12}.

Cold hybrid ion-atom experiments can be used to realize and investigate cold collisions~\cite{CotePRA00,ZipkesNature10,RaviNatCommun12}, controlled chemical reactions~\cite{HallPRL12,SikorskyNC18}, charge and spin transfer dynamics~\cite{RatschbacherNatPhys12,FurstPRA18}, quantum simulation~\cite{BissbortPRL13}, and quantum computation~\cite{DoerkPRA10}. In such systems, diatomic molecular ions can be produced via spontaneous or stimulated charge-transfer radiative association or photoassociation~\cite{TomzaPRA15a,daSilvaNJP2015}, however, only RbCa$^+$~\cite{HallPRL11,HallMP13a}, RbBa$^+$~\cite{HallMP13b}, CaYb$^+$~\cite{RellergertPRL11}, CaBa$^+$~\cite{SullivanPRL12} molecular ions have been observed as products of cold collisions between respective ions and atoms. For higher atomic densities, the three-body processes resulting in the formation of molecular ions additionally play a role~\cite{HartePRL12,KrukowPRL16,WolfScience17}. Cold molecular ions can also be produced by the ionization of ultracold molecules~\cite{SullivanPCCP11,JyothiPRL16,SchmidPRL18} or by sympathetic cooling of molecular ions from higher temperature~\cite{RellergertNature13,HansenNature14}. Molecular ions, which possess additional rotational and vibrational degrees of freedom, can likewise be immersed into ultracold atomic gases opening the way for novel applications in cold controlled ion-atom chemistry~\cite{DeiglmayrPRA12,TomzaPRL15,PuriScience17,KilajNC18,TomzaPCCP17,Dorfler2019,PuriNatChem19,KasPRA19}, precision measurement~\cite{GermannNatPhys14,Cairncross17}, and quantum simulation of many-body physics~\cite{MidyaPRA16}.
In this context, we investigate theoretically the two-body and three-body interactions and chemical reactions in ionic alkali and alkaline-earth diatomic and triatomic systems.

Three-body and many-body nonadditive interactions are important for understanding the emergence of bulk matter properties, crucial across all areas of physics~\cite{IssendorffARPC05}. They have been theoretically investigated in neutral spin-polarized triatomic molecules consisting of alkali-metal atoms~\cite{SoldanPRA03,SoldanPRA08,KlosJCP08,SoldanPRA10,SoldanPRA10b,TomzaPRA13b}, alkali-earth-metal atoms~\cite{KaplanPRA96,KaplanJCP00,KlosJCP08}, and Cu, Zn, Au, Ag atoms~\cite{BravoCPL99,KlosJCP08,DanovichJCTC10}. The nonadditive interactions have also been intensely studied in clusters of ions with small molecules~\cite{CastlemanCR86,BieskeCR00}. On the contrary, the knowledge of nonadditive interactions in metallic molecular ions or ionic clusters is limited~\cite{PavoliniJCP87,SpiegelmannJCP88,KhannaPRL88,JeungCPL90,HeerRMP93,MartinezPRB94,SmartJMS96,LyalinPRA07}.

In this paper, we theoretically investigate the ground-state electronic structure of single-charged molecular ions formed from two or three interacting alkali-metal and alkaline-earth-metal atoms. We calculate ground-state electronic properties of all diatomic $AB^+$ and most of triatomic $A_2B^+$ molecular ions  consisting of Li, Na, K, Rb, Cs, Mg, Ca, Sr, Ba, and Yb atoms using \textit{ab initio} techniques of quantum chemistry. We obtain equilibrium distances, atomization energies, ionization potentials, permanent electric dipole moments, and polarizabilities. A variety of equilibrium geometries for the trimers from linear through isosceles triangular to equilateral triangular are discovered. Furthermore, we evaluate and characterize three-body nonaddtive interactions in these systems at equilibrium geometries. We also provide two-dimensional interaction potential energy surfaces for exemplary non-reactive KRb$^+$+K and Sr$_2$+Rb$^+$ mixtures. We identify possible channels of chemical reactions in ionic two-body: $A^+$+B and $AB^+$, and three-body systems: $A^++AB$, $AB^++A$, and $A_2B^+$, based on the energetics of the reactants. Additionally, we present example calculations of minimum energy paths for the isomerisation reaction of linear alkali-metal trimers in the lowest triplet electronic state between asymmetric $AAB^+$ and symmetric $ABA^+$ arrangements. The present results may be useful for investigating controlled chemical reactions and other applications of molecular ions in modern experiments with cold ion-atom mixtures.

The paper has the following structure. Section~\ref{sec:theory} describes the theoretical methods used in the \textit{ab initio} electronic structure calculations. Section~\ref{sec:results} presents and discusses the results concerning diatomic molecular ions, triatomic molecular ions, and channels of chemical reactions. Section~\ref{sec:summary} summarizes the paper and points to further applications and extensions of the presented results and methodology.

\section{Theoretical Methods}
\label{sec:theory}

In order to obtain potential energy curves (PECs) and surfaces (PESs) within the Born-Oppenheimer  approximation, we adopt the computational scheme successfully applied to the ground-state interactions between polar alkali-metal dimer~\cite{TomzaPRA13b} and between linear polyatomic anions with alkali-metal and alkaline-earth-metal atoms~\cite{TomzaPCCP17}. Thus, to calculate PECs and PESs we employ  the  close-shell or spin-restricted open-shell coupled  cluster  methods  restricted  to  single, double, and noniterative triple excitations, starting from the  restricted  close-shell or open-shell  Hartree-Fock  orbitals,  CCSD(T)~\cite{PurvisJCP82,KnowlesJCP93}. The interaction energies are obtained with the supermolecule method and the basis set superposition error is corrected by using the counterpoise correction~\cite{BoysMP70}
\begin{equation}\label{eq:V}
\begin{split}
V_{AB}&=E_{AB}-E_{A}-E_{B}\,,\\
V_{ABC}&=E_{ABC}-E_{A}-E_{B}-E_{C}\,,\\
V_{AB+C}&=E_{ABC}-E_{AB}-E_{C}\,,\\
\end{split}
\end{equation}
where $V_{AB}$, $V_{ABC}$, and $V_{AB+C}$ are interaction energies between $A$ and $B$; $A$, $B$, and $C$; and $AB$ within the rigid rotor approximation and $C$, respectively. $E_\textrm{ABC}$, $E_\textrm{AB}$, and $E_\textrm{X}$ denote the total energy of trimer, dimer, and monomer computed in a dimer or trimer basis set. 

The three-body nonadditive interatomic interaction in triatomic molecular ions is defined as 
\begin{equation}
V_{3b}=V_{ABC}-V_{AB}-V_{BC}-V_{AC}\,,
\end{equation}
where $V_{ABC}$ is the interaction energy in the three-atom system while $V_{XY}$ are the two-body interactions, all defined by Eq.~\eqref{eq:V} and calculated in the trimer basis set. The effective three-body interatomic interaction emerges as a many-electron quantum effect despite the fact that only the genuine two-body Coulomb interactions between electrons and nuclei are present in the underlying electronic Hamiltonian.

The Li, Na, and Mg atoms are described with the augmented correlation-consistent polarized core-valence quadruple-$\zeta$ quality basis sets (aug-cc-pCVQZ)~\cite{PrascherTCA10}. The scalar relativistic effects in the K, Rb, Cs, Ca, Sr, Ba, and Yb atoms are included by employing the small-core relativistic energy-consistent pseudopotentials (ECP) to replace the inner-shell electrons~\cite{DolgCR12}. The use of the pseudopotentials allows one to model the inner-shell electron density as accurately as the high quality atomic calculations employed to fit the pseudopotentials and to use larger basis sets to describe the valence electrons. The pseudopotentials from the Stuttgart library are used in presented calculations. 
The K, Ca, Rb, Sr, Cs, Ba, and Yb atoms are  described with the ECP10MDF, ECP10MDF, ECP28MDF, ECP28MDF, ECP46MDF, ECP46MDF, and ECP60MDF pseudopotentials~\cite{LimJCP06,DolgTCA98} and the $[11s11p5d3f]$, $[12s12p7d4f2g]$, $[14s14p7d6f1g]$, $[14s11p6d5f4g]$, $[12s11p6d4f2g]$, $[13s12p6d5f4g]$, and $[10s10p9d5f3g]$ basis sets, respectively, obtained by decontracting and augmenting the basis sets suggested in Refs.~\cite{LimJCP06,DolgTCA98}. The used basis sets were developed in Refs.~\cite{TomzaPCCP11,TomzaMP13,TomzaPRA13a,TomzaPRA14}.
The basis sets are additionally augmented in all calculations for diatomic molecular ions by the set of the $[3s3p2d2f1g]$ bond functions~\cite{midbond}.

To find equilibrium interatomic distances for diatomic molecular ions $AB^+$, we calculate full PECs, whereas for triatomic molecular ions $A_2B^+$ we explore two-dimensional PESs around their minima. To this end, we employ two kinds of a PES minimization. First, and for the majority of systems, we assume an isoscales triangular geometry with each of the $A$ atoms bound to the $B$ atom situated on the symmetry axis of the molecular ion. Therefore the PES becomes a two-dimensional function of two coordinates $V(R,\theta)$, where $R$ is the distance between the $B$ atom and each of the $A$ atoms, and $\theta$ is the angle between the two legs of the triangle. When $\theta$ equals 180 degrees, the molecular ion becomes linear and shows a higher symmetry of the $D_{\infty h}$ point group with the general formula $ABA^+$. The second kind of the PES minimisation applies to linear trimers that fall into the $C_{\infty v}$ symmetry group with the general formula $AAB^+$. Then, the PES is a function of two interatomic distances $V(R_{AA},R_{AB})$, where $R_{AA}$ is the distance between two $A$ atoms, and $R_{AB}$ is the distance between the central $A$ atom and $B$~atom. We confirm that found minima are not saddle points in three-dimensional optimizations.

\begin{figure}[tb!]
\begin{center}
\includegraphics[width=\columnwidth]{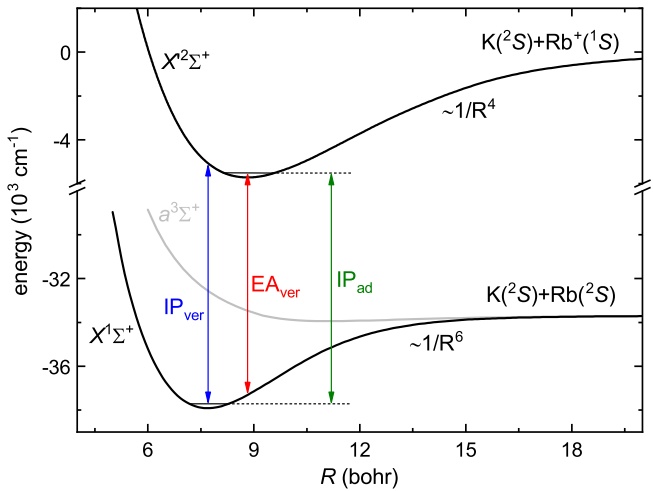}
\end{center}
\caption{Ground-state potential energy curves for the KRb molecule and KRb$^+$ molecular ions together with definitions of the adiabatic and vertical ionization potentials (IP$_\mathrm{ad}$ and IP$_\mathrm{ver}$), and the vertical electron attachment energy (EA$_\mathrm{ver}$).}
\label{fig:scheme}
\end{figure}

The static electric dipole and quadrupole polarizabilities of atoms and the polarizabilities and permanent electric dipole moments of diatomic molecular ions  are calculated with the CCSD(T) and finite field methods. The $z$ axis is chosen along the internuclear axis and is oriented from an atom with a larger ionization potential to an atom with a smaller ionization potential, and the origin is set in the center of mass. The adiabatic and vertical ionization potentials (IP$_\mathrm{ad}$ and IP$_\mathrm{ver}$) and the vertical electron attachment energies (EA$_\mathrm{ver}$) are extracted from energy calculations for diatomic molecular ions $AB^+$ and neutral molecules $AB$ as presented for the exemplary KRb$^+$ molecular ion in Fig.~\ref{fig:scheme}.

To assess the accuracy of the employed \textit{ab initio} methods, Table~\ref{tab:atoms} collects the static electric dipole and quadrupole polarizabilities, ionization potentials, and the lowest $S-P$ excitation energies of alkali-metal and alkaline-earth-metal atoms. Present theoretical values are compared with the most accurate available theoretical or experimental data. The calculated static electric dipole and quadrupole polarizabilities agree with previous data within 0.1-5.7$\,$a.u.~and 9-314$\,$a.u.~that correspond to an error of 0.1-2.2$\,\%$ and 0.6-3.7$\,\%$, respectively. The ionization potentials and the lowest $S-P$ excitation energies coincide with experimental results within 23-255$\,$cm$^{-1}$ and 7-190$\,$cm$^{-1}$ that is 0.05-0.6$\,\%$ and 0.05-1$\,\%$, respectively. Additionally, we compare the available experimental data for the $^1\Sigma^+$ state of all 15 neutral alkali-metal molecules~\cite{ZuchowskiPRA10} with calculated values, and the mean absolute error for the dissociation energy is 74$\,$cm$^{-1}$ (1.6$\,\%$), while for the equilibrium bound length it is 0.013$\,$bohr ($0.18\,\%$). Overall agreement between calculated properties and the most accurate available theoretical or experimental data is very good. This confirms that the employed CCSD(T) method, energy-consistent pseudopotentials, and basis sets properly reproduce correlation energy and include relativistic effects, while being close to convergence in the size of the basis function set. Thus, the used methodology should also provide an accurate description of interaction energies for investigated molecular ions. Based on the above and our previous experience, we estimate the total uncertainty of the calculated PECs for diatomic molecular ions and PESs for triatomic molecular ions at the global minima to be of the order of 100-300$\,$cm$^{-1}$ that corresponds to 2-5\% of the interaction energy. Larger uncertainty may be expected in systems with larger number of valence electrons. The lack of the exact treatment of the triple and higher excitations in the employed CCSD(T) method and the quality of employed energy-consistent pseudopotentials in reproduction of relativistic effects are primary limiting factors.

\begin{table}[tb!]
\caption{Characteristics of alkali-metal and alkaline-earth-metal atoms: the static electric dipole polarizability $\alpha$, the static electric quadrupole polarizability $\beta$, the ionization potential IP, and the lowest $S$--$P$ excitation energy (${}^2S$--${}^2P$ for alkali-metal atoms and ${}^1S$--${}^3P$ for alkaline-earth-metal atoms). Present theoretical values are compared with the most accurate available theoretical or experimental data.\label{tab:atoms}}  
\begin{ruledtabular}
\begin{tabular}{lllll}
Atom & $\alpha\,$(a.u.) & $\beta\,$(a.u.) & IP$\,$(cm$^{-1}$) & $S$--$P$$\,$(cm$^{-1}$)\\
\hline
Li &  164.3 &  1414 & 43464 &  14911 \\
   &  164.2~\cite{MiffreEPJ06} & 1423~\cite{YanPRA96} & 43487~\cite{nist} & 14904~\cite{nist} \\
Na &  166.4 &  1920 & 41217 & 16799 \\
   &  162.7~\cite{EkstromPRA95} & 1895~\cite{KaurPRA15} & 41449~\cite{nist} &  16968~\cite{nist}     \\
K  &  290.8 &  4970 & 34949  & 13022 \\
   &  290.0~\cite{GregoirePRA15}  & 4947~\cite{KaurPRA15} & 35010~\cite{nist} & 13024~\cite{nist}      \\
Rb &  319.5 &  6578 & 33566  & 12686 \\
   &  320.1~\cite{GregoirePRA15} &  6491~\cite{KaurPRA15} & 33691~\cite{nist} &  12737~\cite{nist}     \\
Cs &  395.5 & 10343 & 31331 & 11594 \\
   &  401.2~\cite{GregoirePRA15}   & 10470~\cite{PorsevJCP03} & 31406~\cite{nist} &  11548~\cite{nist}     \\
Mg &   71.8 &   821 & 61466  & 21701 \\
   &   71.3~\cite{PorsevJETP}  &  812~\cite{PorsevJETP} & 61671~\cite{nist} & 21891~\cite{nist}     \\
Ca &  156.9 &  2946 & 49243  & 15190 \\
   &  157.1~\cite{PorsevJETP}  &  3081~\cite{PorsevJETP} & 49306~\cite{nist}  & 15263~\cite{nist}      \\
Sr &  199.2 &  4551 & 45814  & 14639 \\
   &  197.2~\cite{PorsevJETP} &  4630~\cite{PorsevJETP} & 45932~\cite{nist} & 14705~\cite{nist}   \\
Ba &  276.8 &  8586 & 41780 & 13106 \\
   &  273.5~\cite{PorsevJETP} &  8900~\cite{PorsevJETP} & 42035~\cite{nist}  &  13099~\cite{nist}  \\
Yb &  143.5 & 2642 & 50267 & 17635 \\
   &  141.0~\cite{PorsevPRA14} & 2560~\cite{PorsevPRA14} & 50443~\cite{nist} & 18903~\cite{nist} \\
\end{tabular}
\end{ruledtabular}
\end{table}

\begin{table*}[tb!]
\caption{Characteristics of alkali-metal diatomic molecular ions in the $^2\Sigma^+$ electronic ground state: equilibrium interatomic distance $R_e$ (in bohr), well depth $D_e$ (in cm$^{-1}$), harmonic constant $\omega_e$ (in cm$^{-1}$), rotational constant $B_e$ (in cm$^{-1}$), permanent electric dipole moment $d_e$ (in D), perpendicular and parallel components of the static electric dipole polarizability $\alpha_e^\parallel$ and $\alpha_e^\perp$ (in a.u.), vertical electron attachment energy EA$_\mathrm{ver}$ (in cm$^{-1}$), vertical and adiabatic ionization potential of corresponding neutral molecule IP$_\mathrm{ver}$ (in cm$^{-1}$) and IP$_\mathrm{ad}$ (in cm$^{-1}$), and dissociation limit. Previous calculations are cited in the last column.\label{tab:AB+_A}} 
\begin{ruledtabular}
\begin{tabular}{lrrrrrrrrrrrr}
$AB^+$  & $R_e$ & $D_e$ & $\omega_e$ & $B_e$ & $d_e$ & $\alpha_e^\parallel$ & $\alpha_e^\perp$ & EA$_\mathrm{ver}$ & IP$_\mathrm{ver}$ & IP$_\mathrm{ad}$ & Diss. & Previous \\
\hline
 Li$_2^+$ & 5.85 & 10451 & 263 & 0.500 & 0 & 54.4 & 61.3 & 40662 & 42392 & 41521 & Li$^+$+Li & \cite{MagnierCP99,BouzouitaJMS06,JasikEPJ07,MusialMP15,RabliCP17,SchmidPRL18}\\
 LiNa$^+$ & 6.36 & 8047 & 192 & 0.277 & -3.76 & 88.8 & 72.1 & 39623 & 41377 & 40474 &  Na$^+$+Li & \cite{MagnierJPCA01,BerricheJMS03,KhelifiJRLR09,RabliCP18}\\
 LiK$^+$ & 7.28 & 4822 & 154 & 0.191 & -3.59 & 120 & 96.1 & 35537 & 37169 & 36346 & K$^+$+Li & \cite{BerricheJMS05,AldossaryRJP14,MusialaAQC18} \\
 LiRb$^+$ & 7.61 & 4146 & 139 & 0.160 & -4.69 & 136 & 103 & 34621 & 36234 & 35431 & Rb$^+$+Li & \cite{GhanmiIJQC11}\\
 LiCs$^+$ & 8.07 & 3478 & 126 & 0.140 & -4.75 & 156 & 114 & 32893 & 34544 & 33736 & Cs$^+$+Li & \cite{GhanmiJMS06b,KhelifiJAS11} \\
 Na$_2^+$ & 6.81 & 7961 & 120 & 0.113 & 0 & 136 & 82.2 & 38640 & 40375 & 39466 & Na$^+$+Na & \cite{BerricheIJQC13,BewiczMP17}\\
 NaK$^+$ & 7.69 & 4644 & 91.6 & 0.0704 & 0.20 & 176 & 107 & 34806 & 36361 & 35565 & K$^+$+Na & \cite{GhanmiJMS06,MusialaAQC18}\\
 NaRb$^+$ & 8.01 & 3972 & 77.2 & 0.0519 & -2.35 & 194 & 114 & 33928 & 35456 & 34682 & Rb$^+$+Na & \cite{GhanmiJMS07}\\
 NaCs$^+$ & 8.44 & 3226 & 68.5 & 0.0431 & -2.93 & 209 & 124 & 32299 & 33768 & 33046 & Cs$^+$+Na & \cite{GhanmiJMS06b,KorekCJP08}\\
 K$_2^+$ & 8.52 & 6625 & 72.7 & 0.0426 & 0 & 265 & 150 & 32115 & 33416 & 32742 & K$^+$+K & \cite{IlyabaevJCP93,MagnierJQSRT03,BerricheJCMSE08,JraijCJP08,SkupinJPCA17,RabliCPL19}\\
 KRb$^+$ & 8.81 & 5728 & 60.0 & 0.0290 & -3.38 & 303 & 161 & 31488 & 32748 & 32092 & Rb$^+$+K & \cite{KorekIJQC03}\\
 KCs$^+$ & 9.24 & 4626 & 53.5 & 0.0234 & -4.24 & 335 & 179 & 30167 & 31361 & 30746 & Cs$^+$+K & \cite{KorekCJP02} \\
 Rb$_2^+$ & 9.10 & 6151 & 46.0 & 0.0171 & 0 & 347 & 173 & 30863 & 32074 & 31443 & Rb$^+$+Rb & \cite{JraijCP03,AymarJPB03}\\
 RbCs$^+$ & 9.53 & 4952 & 39.8 & 0.0128 & -1.19 & 391 & 193 & 29637 & 30887 & 30185 & Cs$^+$+Rb &\cite{KorekJPB01,AymarJPB03} \\
 Cs$_2^+$ & 9.95 & 5796 & 34.0 & 0.00915 & 0 & 452 & 217 & 28630 & 29699 & 29140 & Cs$^+$+Cs & \cite{AymarJPB03,JraijCP05} \\
\end{tabular}
\end{ruledtabular}
\end{table*}
\begin{table*}[tb!]
\caption{Characteristics of alkaline-earth-metal diatomic molecular ions in the $^2\Sigma^+$ electronic ground state: equilibrium interatomic distance $R_e$ (in bohr), well depth $D_e$ (in cm$^{-1}$), harmonic constant $\omega_e$ (in cm$^{-1}$), rotational constant $B_e$ (in cm$^{-1}$), permanent electric dipole moment $d_e$ (in D), perpendicular and parallel components of the static electric dipole polarizability $\alpha_e^\parallel$ and $\alpha_e^\perp$ (in a.u.), vertical electron attachment energy EA$_\mathrm{ver}$ (in cm$^{-1}$), vertical and adiabatic ionization potential of corresponding neutral molecule IP$_\mathrm{ver}$ (in cm$^{-1}$) and IP$_\mathrm{ad}$ (in cm$^{-1}$), and dissociation limit. Previous calculations are cited in the last column. \label{tab:AB+_AE}} 
\begin{ruledtabular}
\begin{tabular}{lrrrrrrrrrrrr}
$AB^+$ & $R_e$ & $D_e$ & $\omega_e$ & $B_e$ & $d_e$ & $\alpha_e^\parallel$ & $\alpha_e^\perp$ & EA$_\mathrm{ver}$ & IP$_\mathrm{ver}$ & IP$_\mathrm{ad}$ & Diss. & Previous \\
\hline
 Mg$_2^+$ & 5.70 & 10532 & 215 & 0.154 & 0 & 143 & 75.7 & 50233 & 54898 & 51503 & Mg$^+$+Mg & \cite{HuidongMP13,AlharzaliJCPB18} \\
 MgCa$^+$ & 6.53 & 5334 & 153 & 0.0941 & -0.51 & 239 & 122 & 43887 & 45769 & 44549 & Ca$^+$+Mg & \\
 MgSr$^+$ & 6.92 & 4216 & 122 & 0.0667 & -1.37 & 253 & 141 & 41770 & 43234 & 42312 & Sr$^+$+Mg & \\
 MgBa$^+$ & 7.30 & 3569 & 107 & 0.0553 & -2.06 & 269 & 178 & 38722 & 39784 & 39139 & Ba$^+$+Mg & \\
 MgYb$^+$ & 6.88 & 4280 & 118 & 0.0603 & -3.35 & 209 & 106 & 45481 & 48268 & 46482 & Yb$^+$+Mg & \\
 Ca$_2^+$ & 7.23 & 9285 & 130 & 0.0576 & 0 & 293 & 167 & 40425 & 41880 & 40935 & Ca$^+$+Ca & \cite{BanerjeeCPJ12,HuidongMP13}\\
 CaSr$^+$ & 7.58 & 7273 & 105 & 0.0381 & -2.42 & 372 & 186 & 39115 & 40440 & 39594 & Sr$^+$+Ca & \\
 CaBa$^+$ & 7.97 & 5946 & 91.4 & 0.0306 & -3.42 & 412 & 228 & 36795 & 37745 & 37161 & Ba$^+$+Ca & \cite{SullivanPRL12} \\
 CaYb$^+$ & 7.46 & 7585 & 98.7 & 0.0333 & 6.07 & 315 & 155 & 41124 & 44287 & 42180 & Ca$^+$+Yb & \cite{RellergertPRL11} \\
 Sr$_2^+$ & 7.92 & 8575 & 80.2 & 0.0218 & 0 & 366 & 210 & 37849 & 39137 & 38310 & Sr$^+$+Sr & \cite{HuidongMP13}\\
SrBa$^+$ & 8.29 & 6825 & 68.6 & 0.0163 & -1.16 & 460 & 250 & 35911 & 36940 & 36299 & Ba$^+$+Sr & \\
SrYb$^+$ & 7.82 & 5920 & 68.2 & 0.0168 & 4.75 & 341 & 178 & 39535 & 42234 & 40486 & Sr$^+$+Yb & \\
Ba$_2^+$ & 8.67 & 8596 & 59.8 & 0.0116 & 0 & 476 & 290 & 34362 & 35207 & 34690 & Ba$^+$+Ba & \cite{HuidongMP13} \\
BaYb$^+$ & 8.21 & 4826 & 54.8 & 0.0116 & 3.90 & 355 & 220 & 36874 & 39124 & 37726 & Ba$^+$+Yb & \\
Yb$_2^+$ & 7.76 & 7089 & 56.8 & 0.0115 & 0 & 330 & 142 & 42410 & 46143 & 43645 & Yb$^+$+Yb & \cite{ZhangPRA09} \\
\end{tabular}
\end{ruledtabular}
\end{table*}

\begin{table*}[tb!]
\caption{Characteristics of alkali-metal--alkaline-earth-metal diatomic molecular ions in the $^1\Sigma^+$ electronic ground state: equilibrium interatomic distance $R_e$ (in bohr), well depth $D_e$ (in cm$^{-1}$), harmonic constant $\omega_e$ (in cm$^{-1}$), rotational constant $B_e$ (in cm$^{-1}$), permanent electric dipole moment $d_e$ (in D), perpendicular and parallel components of the static electric dipole polarizability $\alpha_e^\parallel$ and $\alpha_e^\perp$ (in a.u.), vertical electron attachment energy EA$_\mathrm{ver}$ (in cm$^{-1}$), vertical and adiabatic ionization potential of corresponding neutral molecule IP$_\mathrm{ver}$ (in cm$^{-1}$) and IP$_\mathrm{ad}$ (in cm$^{-1}$), and dissociation limit. Previous calculations are cited in the last column.\label{tab:AB+_AAE}} 
\begin{ruledtabular}
\begin{tabular}{lrrrrrrrrrrrr}
$AB^+$  & $R_e$ & $D_e$& $\omega_e$ & $B_e$ & $d_e$ & $\alpha_e^\parallel$ & $\alpha_e^\perp\,$ & EA$_\mathrm{ver}$ & IP$_\mathrm{ver}$ & IP$_\mathrm{ad}$ & Diss. & Previous \\
\hline
LiMg$^+$ & 5.48 & 6628 & 266 & 0.369 & 5.45 & 75.9 & 56.0 & 38159 & 38483 & 38291 & Li$^+$+Mg & \cite{ElOualhaziJCPA16,BalaMP19}\\
LiCa$^+$ & 6.16 & 9941 & 246 & 0.266 & 4.54 & 156 & 105 & 35844 & 35985 & 35911 & Li$^+$+Ca & \cite{HabliMP16,SaitoPRA17,BalaMP19} \\ 
LiSr$^+$ & 6.45 & 11126 & 230 & 0.223 & 5.04 & 196 & 124 & 34619 & 34786 & 34699 & Li$^+$+Sr & \cite{AymarJCP11,JellaliMP16} \\
LiBa$^+$ & 6.71 & 11674 & 225 & 0.200 & -4.06 & 271 & 147 & 33407 & 33472 & 33435 & Ba$^+$+Li & \\
LiYb$^+$ & 6.29 & 8719 & 225 & 0.225 & 7.28 & 139 & 92.2 & 34848 & 36781 & 35569 & Li$^+$+Yb & \cite{TomzaPRA15a,daSilvaNJP2015} \\
NaMg$^+$ & 6.09 & 4517 & 151 & 0.138 & 3.20 & 88.3 & 60.2 & 37590 & 38038 & 37755 & Na$^+$+Mg & \\
NaCa$^+$ & 6.70 & 7336 & 137 & 0.0918 & 2.54 & 182 & 114 & 35542 & 35701 & 35604 & Na$^+$+Ca & \cite{MakarovPRA03,GacesaPRA16,JellaliJQSRT18} \\
NaSr$^+$ & 6.97 & 8418 & 121 & 0.0680 & 4.34 & 226 & 135 & 34446 & 34668 & 34533 & Na$^+$+Sr & \cite{AymarJCP11,AyedEPJD17,BellaouiniEPJ18} \\ 
NaBa$^+$ & 7.24 & 10318 & 115 & 0.0582 & 3.93 & 307 & 164 & 33110 & 33220 & 33169 & Na$^+$+Ba & \\
NaYb$^+$ & 6.82 & 6386 & 114 & 0.0637 & 7.75 & 167 & 100 & 34877 & 36927 & 35591 & Na$^+$+Yb & \\
KMg$^+$ & 7.07 & 2614 & 102 & 0.0811 & 3.24 & 92.9 & 66.8 & 32917 & 33242 & 33045 & K$^+$+Mg & \cite{FarjallahJPB19}\\ 
KCa$^+$ & 7.71 & 4281 & 93.2 & 0.0513 & 3.37 & 192 & 129 & 31794 & 31946 & 31854 & K$^+$+Ca & \\
KSr$^+$ & 7.99 & 4950 & 79.5 & 0.0350 & 6.27 & 241 & 156 & 31104 & 31326 & 31191 & K$^+$+Sr & \cite{AymarJCP11,SouissiCP17}\\ 
KBa$^+$ & 8.30 & 6170 & 75.8 & 0.0288 & 6.69 & 325 & 197 & 30211 & 30333 & 30259 & K$^+$+Ba & \\
KYb$^+$ & 7.81 & 3688 & 71.4 & 0.0310 & 10.3 & 176 & 113 & 31192 & 32674 & 31759 & K$^+$+Yb & \\
RbMg$^+$ & 7.41 & 2237 & 84.1 & 0.0586 & 0.78 & 99.2 & 70.4 & 31933 & 32227 & 32051 & Rb$^+$+Mg & \\
RbCa$^+$ & 8.06 & 3666 & 73.9 & 0.0341 & 0.39 & 202 & 134 & 30997 & 31140 & 31054 & Rb$^+$+Ca & \cite{TacconiPCCP11,daSilvaNJP2015} \\
RbSr$^+$ & 8.34 & 4247 & 59.0 & 0.0201 & 3.41 & 253 & 162 & 30397 & 30608 & 30479 & Rb$^+$+Sr & \cite{AymarJCP11,daSilvaNJP2015} \\
RbBa$^+$ & 8.65 & 5314 & 54.3 & 0.0153 & 4.33 & 339 & 206 & 29611 & 29737 & 29661 & Rb$^+$+Ba & \cite{KnechtJPB10,KrychPRA11,daSilvaNJP2015} \\
RbYb$^+$ & 8.15 & 3152 & 49.6 & 0.0159 & 8.41 & 184 & 118 & 30436 & 31745 & 30949 & Rb$^+$+Yb & \cite{LambPRA12,SayfutyarovaPRA13} \\
CsMg$^+$ & 7.85 & 1861 & 73.2 & 0.0481 & -0.03 & 109 & 76.6 & 30022 & 30262 & 30122 & Cs$^+$+Mg & \\
CsCa$^+$ & 8.53 & 3017 & 63.2 & 0.0270 & -0.63 & 213 & 141 & 29336 & 29448 & 29382 & Cs$^+$+Ca & \\
CsSr$^+$ & 8.81 & 3484 & 48.7 & 0.0147 & 2.16 & 264 & 170 & 28850 & 29023 & 28919 & Cs$^+$+Sr & \cite{AymarJCP11} \\
CsBa$^+$ & 9.14 & 4350 & 44.0 & 0.0106 & 3.28 & 351 & 217 & 28228 & 28341 & 28273 & Cs$^+$+Ba & \\
CsYb$^+$ & 8.61 & 2591 & 39.3 & 0.0108 & 7.37 & 194 & 125 & 28755 & 29867 & 29213 & Cs$^+$+Yb & \\
\end{tabular}
\end{ruledtabular}
\end{table*}

All electronic structure calculations are performed with the \textsc{Molpro} package of \textit{ab initio} programs \cite{Molpro}. The isosurfaces of electronic density for selected trimers are generated with the Gaussian software~\cite{gaussian}.

\section{Results and Discussion}
\label{sec:results}
\subsection{Diatomic molecular ions $AB^+$}

The electronic ground state of diatomic molecular ions $AB^+$ composed of either two alkali-metal or two alkaline-earth-metal atoms is of an open-shell doublet $^2\Sigma^+$ symmetry, which becomes $^2\Sigma_g^+$ for homonuclear alkali-metal ions and $^2\Sigma_u^+$ for homonuclear alkaline-earth-metal ions. Diatomic molecular ions containing one alkali-metal atom and one alkaline-earth-metal atom have a closed-shell singlet $^1\Sigma^+$ symmetry. All considered diatomic molecular ions, with the exception of LiBa$^+$, are described well by single-reference wave functions at all internuclear distances. The closed-shell LiBa$^+$ molecular ion dissociates into two open-shell atoms.

The electronic structure of diatomic molecular ions have previously been studied for several atomic combinations, including:
Li$_2^+$~\cite{MagnierCP99,BouzouitaJMS06,JasikEPJ07,MusialMP15,RabliCP17,SchmidPRL18},
Na$_2^+$~\cite{BerricheIJQC13,BewiczMP17},
K$_2^+$~\cite{IlyabaevJCP93,MagnierJQSRT03,BerricheJCMSE08,JraijCJP08,SkupinJPCA17,RabliCPL19},
Rb$_2^+$~\cite{JraijCP03,AymarJPB03},
Cs$_2^+$~\cite{AymarJPB03,JraijCP05},
LiNa$^+$~\cite{MagnierJPCA01,BerricheJMS03,KhelifiJRLR09,RabliCP18},
LiK$^+$~\cite{BerricheJMS05,AldossaryRJP14,MusialaAQC18},
LiRb$^+$~\cite{GhanmiIJQC11},
LiCs$^+$~\cite{GhanmiJMS06b,KhelifiJAS11},
NaK$^+$~\cite{GhanmiJMS06,MusialaAQC18},
NaRb$^+$~\cite{GhanmiJMS07},
NaCs$^+$~\cite{GhanmiJMS06b,KorekCJP08},
KRb$^+$~\cite{KorekIJQC03},
KCs$^+$~\cite{KorekCJP02},
RbCs$^+$~\cite{KorekJPB01,AymarJPB03},
Mg$_2^+$~\cite{HuidongMP13,AlharzaliJCPB18},
Ca$_2^+$~\cite{BanerjeeCPJ12,HuidongMP13},
CaBa$^+$~\cite{SullivanPRL12},
CaYb$^+$~\cite{RellergertPRL11},
Sr$_2^+$~\cite{HuidongMP13},
Ba$_2^+$~\cite{HuidongMP13},
Yb$_2^+$~\cite{ZhangPRA09},
LiMg$^+$\cite{ElOualhaziJCPA16,BalaMP19},
KMg$^+$~\cite{FarjallahJPB19},
LiCa$^+$\cite{HabliMP16,SaitoPRA17,BalaMP19},
NaCa$^+$~\cite{MakarovPRA03,GacesaPRA16,JellaliJQSRT18},
RbCa$^+$~\cite{TacconiPCCP11,daSilvaNJP2015},
LiSr$^+$~\cite{AymarJCP11,JellaliMP16},
NaSr$^+$~\cite{AymarJCP11,AyedEPJD17,BellaouiniEPJ18},
KSr$^+$~\cite{AymarJCP11,SouissiCP17},
RbSr$^+$~\cite{AymarJCP11,daSilvaNJP2015},
CsSr$^+$~\cite{AymarJCP11},
RbBa$^+$~\cite{KnechtJPB10,KrychPRA11,daSilvaNJP2015},
LiYb$^+$~\cite{TomzaPRA15a,daSilvaNJP2015},
RbYb$^+$~\cite{LambPRA12,SayfutyarovaPRA13,daSilvaNJP2015}.

\begin{figure*}[tb!]
\begin{center}
\includegraphics[width=0.95\textwidth]{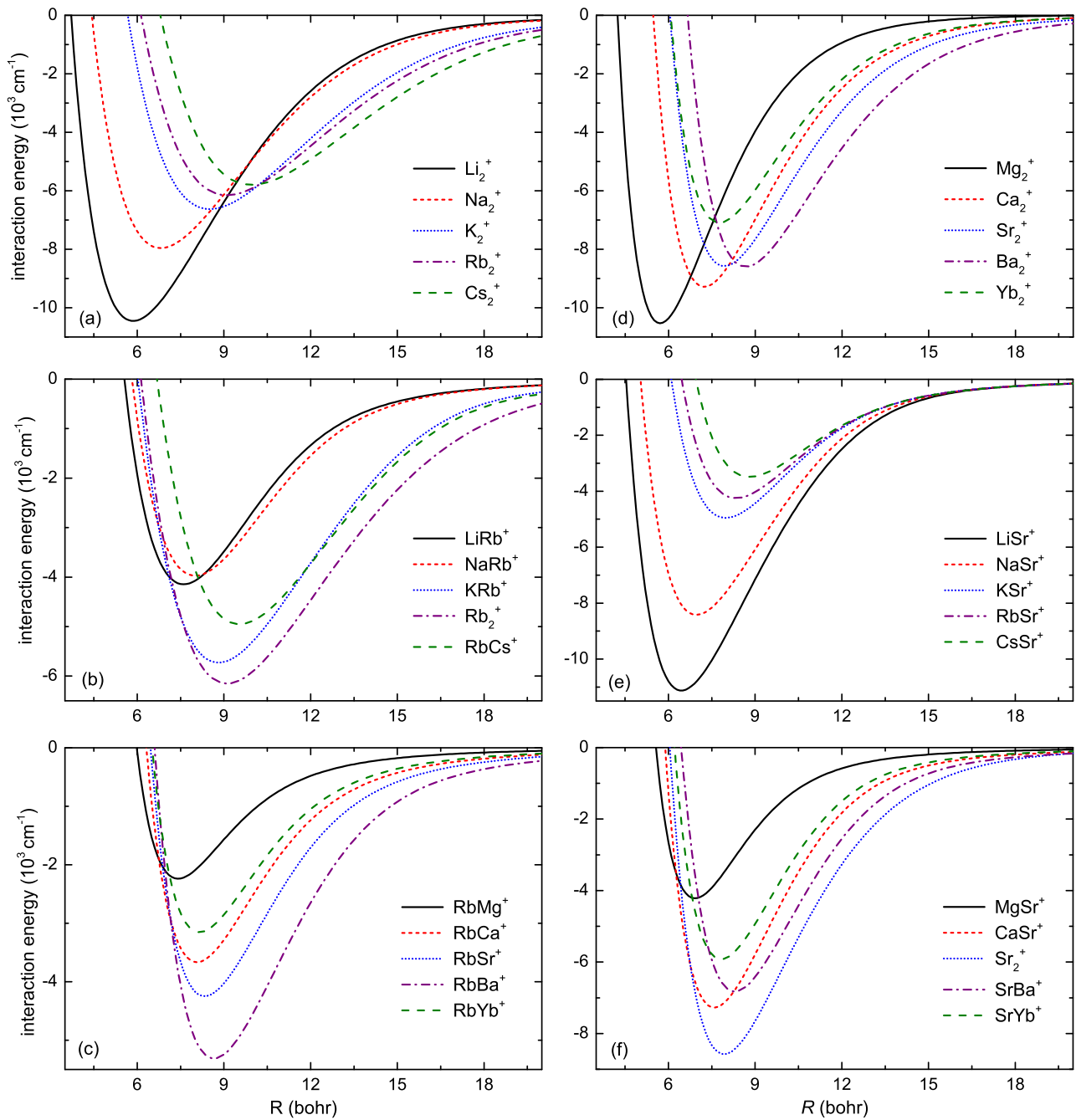}
\end{center}
\caption{Ground-state potential energy curves for selected diatomic molecular ions: (a) homonuclear alkali-metal dimers, (b) heteronuclear alkali-metal dimers containing Rb atom, (c) heteronuclear alkali-metal--alkaline-earth-metal dimers containing Rb atom, (d) homonuclear alkaline-earth-metal dimers, (e) heteronuclear alkali-metal--alkaline-earth-metal dimers containing Sr atom, (f) heteronuclear alkaline-earth-metal dimers containing Sr atom.}
\label{fig:AB+}
\end{figure*}

\begin{figure}[tb!]
\begin{center}
\includegraphics[width=\columnwidth]{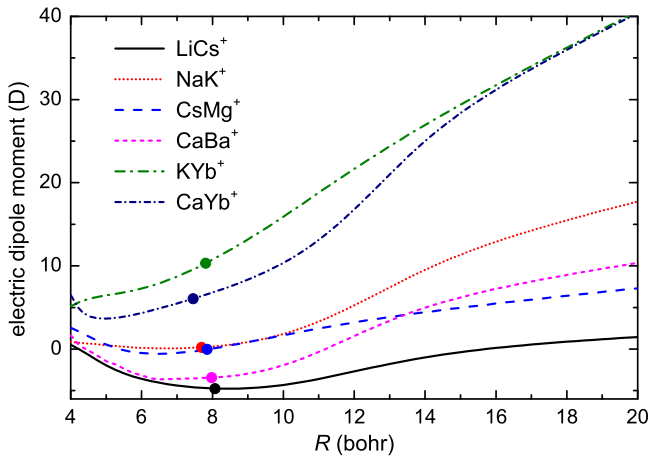}
\end{center}
\caption{Permanent electric dipole moments for selected diatomic molecular ions. The $z$ axis is oriented from the atom with a larger ionization potential to the atom with a smaller ionization potential and the origin is in the center of mass. The points indicate values for equilibrium distances.}
\label{fig:dip}
\end{figure}

Here, for completeness and comparison, and to analyze trends, we calculate the electronic ground-state properties of all 55 diatomic molecular ions consisting of Li, Na, K, Rb, Cs, Mg, Ca, Sr, Ba, and Yb atoms. Among investigated homo- and heteronuclear molecular ions, 15 species consist of two alkali-metal atoms, 15 species consist of two alkaline-earth-metal atoms, and 25 species consist of one alkali-metal atom and one alkaline-earth-metal atom. The calculated properties include: ground-state potential energy curves, equilibrium interatomic distances $R_e$, well depths $D_e$, harmonic constants $\omega_e$, rotational constants $B_e$, permanent electric dipole moments $d_e$, perpendicular and parallel components of the static electric dipole polarizabilities $\alpha_e^\parallel$ and $\alpha_e^\perp$, vertical electron attachment energies EA, and vertical and adiabatic ionization potentials of corresponding neutral molecules IP$_\text{ver}$ and IP$_\text{ad}$. They are collected in Table~\ref{tab:AB+_A} for alkali-metal molecular ions, in Table~\ref{tab:AB+_AE} for alkaline-earth-metal molecular ions, and in Table~\ref{tab:AB+_AAE} for mixed alkali-metal--alkaline-earth-metal molecular ions. To calculate harmonic and rotational constants, atomic masses of the most abundant isotopes are assumed. Additionally PECs and permanent electric dipole moments for selected systems are presented in Fig.~\ref{fig:AB+} and Fig.~\ref{fig:dip}, respectively. Full potential energy curves, permanent electric dipole moments, and electric dipole polarizabilities as a function of interatomic distance are available for all investigated systems from the authors upon request.

Ion-atom interactions are dominated by the induction component, which can be understood in terms of the interaction of the charge of an ion with
the electronic cloud of a neutral partner~\cite{TomzaRMP19}. Therefore, PECs for all investigated diatomic molecular ions have the average of the well depths around 6000$\,$cm$^{-1}$ and are deeper than PECs for corresponding neutral molecules. Among alkali-metal dimers, the Li$_2^+$ molecular ion is the most strongly bound with $D_e=10451\,$cm$^{-1}$, while the NaCs$^+$ molecular ion is the most weakly bound with $D_e=3226\,$cm$^{-1}$. Among alkaline-earth-metal dimers, the Mg$_2^+$ molecular ion is the most strongly bound with $D_e=10532\,$cm$^{-1}$, while the MgBa$^+$ molecular ion is the most weakly bound with $D_e=3569\,$cm$^{-1}$. Among mixed alkali-metal--alkaline-earth-metal dimers the range of binding energies is the largest. For this class of compounds, the LiBa$^+$ molecular ion is the most strongly bound with $D_e=11674\,$cm$^{-1}$, while the CsMg$^+$ molecular ion is the most weakly bound with $D_e=1861\,$cm$^{-1}$. For all considered groups of molecular ions, larger binding energies are mostly correlated with smaller equilibrium distances.  All equilibrium interatomic distances are between 5.45$\,$bohr for LiMg$^+$ and 9.95$\,$bohr for Cs$_2^+$ with the average around 7.5$\,$bohr.

In general, all homonuclear diatomic molecular ions are more strongly bound than related heteronuclear ones due to the more efficient charge delocalization in homonuclear ions. Within homonuclear compounds, the lightest elements have the largest binding energies (see Fig.~\ref{fig:AB+}(a,d)). If the charge is localized at one of the atoms (the one with a smaller ionization potential), then the strength of the interaction is correlated with the polarizability of the neutral partner (see Fig.~\ref{fig:AB+}(b,c) where the charge is mostly localized at Rb, especially at larger internuclear distances, except RbCs$^+$). Within mixed alkali-metal--alkaline-earth-metal molecular ions, with exception of LiBa$^+$, the positive charge is localized at the alkali-metal atom due to its much lower ionization potential (electronegativity) (see Table.~\ref{tab:atoms}). Thus, in the series of molecular ions consisting of the Rb ion and alkaline-earth-metal atom, the dissociation energy increases visibly with increasing polarizability of the alkaline-earth-metal atom and is the largest for the RbBa$^+$ molecular ion while the lowest for the RbMg$^+$ one (see Fig.~\ref{fig:AB+}(c)). In contrast, in the series of molecular ions consisting of the Sr atom and alkali-metal ion, the long-range part of the interaction is determined by the polarizability of the Sr atom and is similar for all compounds. Additionally, a clear trend of an increasing binding energy with the decreasing atomic size is visible for those molecular ions, that indicates a certain degree of covalent bonding (see Fig.~\ref{fig:AB+}(e)).

The permanent electric dipole moments are calculated with respect to the center of mass, which is a natural choice for studying the ro-vibrational dynamics. Exemplary results for selected diatomic molecular ions are presented in Fig.~\ref{fig:dip}. Their absolute values increase with increasing internuclear distance and asymptotically approach the limiting cases $d(R)\approx\frac{\mu}{m_{A^+}}R$, where the charge is completely localized on the ion $A^+$ corresponding to the atom with the smaller ionization potential ($m_{A^+}$ is the mass of the $A^+$ ion and $\mu$ is the reduced mass of the ion-atom system). This behavior is typical for heteronuclear molecular ions and implies that even molecular ions in very weakly bound states have effectively a significant permanent electric dipole moment in contrast to neutral molecules~\cite{TomzaPRA15b}. The differences between calculated values and limiting cases are the interaction-induced variations of the permanent electric dipole moments or, in other words, they describe the degree of charge transfer and delocalization. For most of alkali-metal and alkaline-earth-metal dimers, the calculated permanent dipole moments have a negative sign for the equilibrium distance. This means that the charge is significantly delocalized and transferred from the ion with smaller ionization potential to the second atom due to chemical bonding and orbital mixing at equilibrium distances. For alkali-metal--alkaline-earth-metal molecular ions, the sign of the permanent dipole moments is mostly positive indicating weaker charge transfer and delocalization. In general, the calculated permanent electric dipole moments take values between zero and 10$\,$D for KYb$^+$. Large permanent electric dipole moments may be useful for control with electromagnetic fields or dipolar interactions. Similarly, calculated molecular polarizabilities describe interactions with laser field and may give information about anisotropy of intermolecular interactions~\cite{DeiglmayrJCP08}. 

The ionization potentials for all molecular ions $AB^+$ are smaller than the smallest ionization potential among related constituent atoms $A$ and $B$, because PECs for molecular ions $AB^+$ are deeper than for neutral molecules $AB$ (cf.~Fig.~\ref{fig:scheme}). At the same time the adiabatic and vertical ionization potentials of neutral molecules $AB$, and the vertical electron attachment energies to molecular ions $AB^+$ are very similar, because the equilibrium distances for molecular ions and parent neutral molecules are similar.

\begin{table*}[tb!]
\caption{Characteristics of alkali-metal triatomic molecular ions $A_2B^+$ in the singlet $^1A_1$ electronic ground state. All molecular ions have an isosceles triangular equilibrium geometry within the $C_{2v}$ point group. Consecutive columns list: equilibrium angle between the two even legs of the triangle $\alpha_e$, equilibrium leg length $R_{AB}$, well depth of the triatomic ion $D_e$, additive two-body part of the binding energy $D_{2b}$, nonadditive three-body part of the binding energy $D_{3b}$, and dissociation limit. \label{tab:ABC+_AL_S}} 
\begin{ruledtabular}
\begin{tabular}{lrrrrrr}
$A_2B^+$ & $\alpha_e\,$(degrees) & $R_{AB}\,$(bohr) & $D_e\,$(cm$^{-1}$) & $D_{2b}\,$(cm$^{-1}$) & $D_{3b}\,$(cm$^{-1}$) & Diss. \\
\hline
Li$_3^+$ & 60.0 & 5.62 & 23610 & 28799 & -5190 & Li$^+$+Li+Li \\
Li$_2$Na$^+$ & 53.8 & 6.14 & 20281 & 24119 & -3838 & Na$^+$+Li+Li \\
Li$_2$K$^+$ & 43.6 & 7.21 & 15583 & 17975 & -2391 & K$^+$+Li+Li \\
Li$_2$Rb$^+$ & 41.1 & 7.58 & 14607 & 16666 & -2059 & Rb$^+$+Li+Li \\
Li$_2$Cs$^+$ & 38.0 & 8.08 & 13514 & 15371 & -1857 & Cs$^+$+Li+Li \\
Na$_2$Li$^+$ & 65.7 & 6.08 & 26004 & 22534 & 3470 & Na$^+$+Na+Li\\
Na$_3^+$ & 60.0 & 6.53 & 18317 & 21318 & -3001 & Na$^+$+Na+Na \\
Na$_2$K$^+$ & 48.7 & 7.58 & 13435 & 15070 & -1636 & K$^+$+Na+Na \\
Na$_2$Rb$^+$ & 46.1 & 7.92 & 12417 & 13771 & -1354 & Rb$^+$+Na+Na \\
Na$_2$Cs$^+$ & 42.4 & 8.45 & 11240 & 12344 & -1104 & Cs$^+$+Na+Na\\
K$_2$Li$^+$ & 82.7 & 6.79 & 15305 & 17111 & -1806 & K$^+$+K+Li\\
K$_2$Na$^+$ & 75.6 & 7.19 & 14263 & 16037 & -1773 & K$^+$+K+Na\\
K$_3^+$ & 60.0 & 8.25 & 14425 & 17153 & -2728 & K$^+$+K+K\\
K$_2$Rb$^+$ & 56.5 & 8.60 & 13147 & 15468 & -2321 & Rb$^+$+K+K \\
K$_2$Cs$^+$ & 52.0 & 9.13 & 11495 & 13394 & -1898 & Cs$^+$+K+K \\
Rb$_2$Li$^+$ & 89.1 & 7.05 & 14389 & 15588 & -1199 & Rb$^+$+Rb+Li \\
Rb$_2$Na$^+$ & 81.0 & 7.44 & 13318 & 14665 & -1347 & Rb$^+$+Rb+Na \\
Rb$_2$K$^+$ & 63.8 & 8.49 & 13217 & 15617 & -2400 & Rb$^+$+Rb+K \\
Rb$_3^+$ & 60.0 & 8.85 & 13258 & 15779 & -2521 & Rb$^+$+Rb+Rb \\
Rb$_2$Cs$^+$ & 54.6 & 9.39 & 11502 & 13574 & -2072 & Cs$^+$+Rb+Rb \\
Cs$_2$Li$^+$ & 101.2 & 7.39 & 13793 & 14167 & -373 & Cs$^+$+Cs+Li  \\
Cs$_2$Na$^+$ & 91.0 & 7.76 & 12590 & 13202 & -612 & Cs$^+$+Cs+Na \\
Cs$_2$K$^+$ & 70.4 & 8.81 & 11937 & 14047 & -2110 & Cs$^+$+Cs+K \\
Cs$_2$Rb$^+$ & 66.0 & 9.17 & 11868 & 14143 & -2275 & Cs$^+$+Cs+Rb \\
Cs$_3^+$ & 60.0 & 9.73 & 12118 & 14733 & -2615 & Cs$^+$+Cs+Cs \\
\end{tabular}
\end{ruledtabular}
\end{table*}

Calculated PECs' parameters agree well with results obtained previously for selected systems by other authors, including spectroscopic measurements and calculations using different electronic structure methods such as effective large-core pseudopotentials with core-polarization potentials. Absolute deviations of potentials' well depths between previous and our results in most cases are in the range of 100-300$\,$cm$^{-1}$, within our estimated error bars (see e.g.~\cite{MakarovPRA03,JraijCP03,AymarJCP11,KrychPRA11,GhanmiIJQC11,TacconiPCCP11,LambPRA12,SayfutyarovaPRA13,MusialMP15,daSilvaNJP2015,JellaliMP16} for representative examples). The equilibrium distances agree within 0.05-0.2$\,$bohr. Calculations with large-core pseudopotentials~\cite{AymarJPB03,AymarJCP11,daSilvaNJP2015,HabliMP16,JellaliMP16,SouissiCP17,JellaliJQSRT18,BellaouiniEPJ18} have a tendency to give smaller equilibrium distances but the overall agreement between calculations with small-core and large-core pseudopotentials is good and cross-validates both approaches. For the readability of calculated characteristics collected in Tables~\ref{tab:AB+_A}-\ref{tab:AB+_AAE} and due to space limitations, we do not present detailed comparison of all existing calculations for diatomic molecular ions with our calculations but cite them. Detailed comparison is presented in Supplemental Material.

\subsection{Triatomic molecular ions $A_2B^+$}

Assuming an isosceles triangular geometry, triatomic heteronuclear molecular ions $A_2B^+$ composed of three alkali-metal atoms in the electronic ground state have a closed-shell $^1A_1$ symmetry within the C$_{2v}$ point group. Homonuclear trimers $A_3^+$ of alkali-metal atoms additionally show three-fold rotational symmetry. The lowest energetic triplet state of the alkali-metal trimers have a $^3B_2$ symmetry. The ground state of alkaline-earth-metal trimers has a doublet multiplicity and can be either in a $^2A_1$ or $^2B_2$ representation of the $C_{2v}$ symmetry group. Homonuclear alkaline-earth-metal trimers do not show three-fold symmetry as opposed to homonuclear alkali-metal triatomic ions.

\begin{table*}[tb!p]
\caption{Characteristics of alkali-metal triatomic molecular ions $AAB^+$ and $ABA^+$ in the lowest triplet $^3B_2$ electronic state which coreduces to the $^3\Sigma^+$ and $^3\Sigma^+_u$ symmetries at a linear equilibrium geometry within the $C_{\infty v}$ and $D_{\infty h}$ point groups, respectively. Consecutive columns list: equilibrium distance between the first and the second atom $R_{12}$, equilibrium distance between the second and the third atom $R_{23}$, well depth of the triatomic ion $D_e$, additive two-body part of the binding energy $D_{2b}$, nonadditive three-body part of the binding energy~$D_{3b}$, and dissociation limit.\label{tab:ABC+_AL_T}} 
\begin{ruledtabular}
\begin{tabular}{lrrrrrr}
$ABC^+$  & $R_{12}\,$(bohr) & $R_{23}\,$(bohr) & $D_e\,$(cm$^{-1}$) & $D_{2b}\,$(cm$^{-1}$) & $D_{3b}\,$(cm$^{-1}$) & Diss.\\
\hline
LiLiLi$^+$ & 5.90 & 5.90 & 16856 & 21014 & -4158 & Li$^+$+Li+Li \\
LiNaLi$^+$ & 6.54 & 6.54 & 12843 & 16117 & -3274 & Na$^+$+Li+Li \\
LiLiNa$^+$ & 5.95 & 6.43 & 14173 & 9331 & 4842 & Na$^+$+Li+Li \\
LiKLi$^+$ & 7.46 & 7.46 & 8146 & 9644 & -1498 & K$^+$+Li+Li \\
LiLiK$^+$ & 6.02 & 7.29 & 9669 & 5402 & 4267 & K$^+$+Li+Li \\
LiRbLi$^+$ & 7.80 & 7.80 & 7033 & 8282 & -1249 & Rb$^+$+Li+Li \\
LiLiRb$^+$ & 6.05 & 7.59 & 8650 & 4644 & 4006 & Rb$^+$+Li+Li \\
LiCsLi$^+$ & 8.25 & 8.25 & 5836 & 6947 & -1111 & Cs$^+$+Li+Li \\
LiLiCs$^+$ & 6.09 & 8.01 & 7409 & 3926 & 3483 & Cs$^+$+Li+Li \\
NaLiNa$^+$ & 6.44 & 6.44 & 13712 & 9608 & 4104 & Na$^+$+Na+Li \\
NaNaLi$^+$ & 6.98 & 6.56 & 12539 & 8481 & 4058 & Na$^+$+Na+Li \\
NaNaNa$^+$ & 6.99 & 6.99 & 12270 & 15938 & -3668 & Na$^+$+Na+Na \\
NaKNa$^+$ & 7.86 & 7.86 & 7742 & 9292 & -1550 & K$^+$+Na+Na \\
NaNaK$^+$ & 7.13 & 7.78 & 7770 & 4843 & 2927 & K$^+$+Na+Na \\
NaRbNa$^+$ & 8.19 & 8.19 & 6677 & 12856 & -6179 & Rb$^+$+Na+Na \\
NaNaRb$^+$ & 7.18 & 8.07 & 6794 & 4109 & 2686 & Rb$^+$+Na+Na \\
NaCsNa$^+$ & 8.63 & 8.63 & 5512 & 6443 & -932 & Cs$^+$+Na+Na \\
NaNaCs$^+$ & 7.28 & 8.47 & 5600 & 3290 & 2310 & Cs$^+$+Na+Na \\
KLiK$^+$ & 7.35 & 7.35 & 10657 & 6732 & 3925 & K$^+$+K+Li \\
KKLi$^+$ & 8.63 & 7.56 & 9288 & 6843 & 2445 & K$^+$+K+Li \\
KNaK$^+$ & 7.88 & 7.88 & 9151 & 5876 & 3275 & K$^+$+K+Na \\
KKNa$^+$ & 8.63 & 7.96 & 9096 & 6663 & 2433 & K$^+$+K+Na \\
KKK$^+$ & 8.72 & 8.72 & 10169 & 13256 & -3087 & K$^+$+K+K \\
KRbK$^+$ & 9.03 & 9.03 & 8917 & 11451 & -2533 & Rb$^+$+K+K \\
KKRb$^+$ & 8.74 & 8.98 & 9066 & 6139 & 2927 & Rb$^+$+K+K \\
KCsK$^+$ & 9.47 & 9.47 & 7377 & 9242 & -1864 & Cs$^+$+K+K \\
KKCs$^+$ & 8.80 & 9.37 & 7520 & 4933 & 2587 & Cs$^+$+K+K \\
RbLiRb$^+$ & 7.66 & 7.66 & 9818 & 6001 & 3817 & Rb$^+$+Rb+Li \\
RbRbLi$^+$ & 9.23 & 7.94 & 8350 & 6349 & 2001 & Rb$^+$+Rb+Li \\
RbNaRb$^+$ & 8.19 & 8.19 & 8346 & 5198 & 3148 & Rb$^+$+Rb+Na \\
RbRbNa$^+$ & 9.23 & 8.33 & 8186 & 6182 & 2004 & Rb$^+$+Rb+Na \\
RbKRb$^+$ & 9.01 & 9.01 & 9299 & 6500 & 2799 & Rb$^+$+Rb+K \\
RbRbK$^+$ & 9.30 & 9.06 & 9140 & 6471 & 2669 & Rb$^+$+Rb+K \\
RbRbRb$^+$ & 9.33 & 9.33 & 9342 & 12300 & -2957 & Rb$^+$+Rb+Rb \\
RbCsRb$^+$ & 9.77 & 9.77 & 7744 & 9892 & -2148 & Cs$^+$+Rb+Rb \\
RbRbCs$^+$ & 9.39 & 9.71 & 7725 & 5251 & 2474 & Cs$^+$+Rb+Rb \\
CsLiCs$^+$ & 8.09 & 8.09 & 9114 & 5475 & 3639 & Cs$^+$+Cs+Li \\
CsCsLi$^+$ & 10.09 & 8.43 & 7518 & 6036 & 1482 & Cs$^+$+Cs+Li \\
CsNaCs$^+$ & 8.63 & 8.63 & 7589 & 4599 & 2990 & Cs$^+$+Cs+Na \\
CsCsNa$^+$ & 10.08 & 8.83 & 7378 & 5867 & 1511 & Cs$^+$+Cs+Na \\
CsKCs$^+$ & 9.45 & 9.45 & 8292 & 5531 & 2761 & Cs$^+$+Cs+K \\
CsCsK$^+$ & 10.14 & 9.56 & 8101 & 6039 & 2062 & Cs$^+$+Cs+K \\
CsRbCs$^+$ & 9.77 & 9.77 & 8259 & 5698 & 2561 & Cs$^+$+Cs+Rb \\
CsCsRb$^+$ & 10.16 & 9.82 & 8257 & 6061 & 2196 & Cs$^+$+Cs+Rb \\
CsCsCs$^+$ & 10.22 & 10.22 & 8685 & 11576 & -2891 & Cs$^+$+Cs+Cs \\
\end{tabular}
\end{ruledtabular}
\end{table*}

\begin{table*}[tb]
\caption{Characteristics of symmetric forms of alkaline-earth-metal triatomic molecular ions $ABA^+$ in the lowest doublet $^2A_1$ electronic state within the $C_{2v}$ point group which coreduces to the $^3\Sigma^+_g$ symmetry at a linear equilibrium geometry within $D_{\infty h}$ point group. Calculated potential energy surfaces for most of the molecular ions have two local minima: one at the linear or obtuse angle isosceles triangular geometry (below both listed as linear) and another at the acute angle isosceles triangular geometry (below listed as triangular). Characteristics for both minima are included. Consecutive columns list: equilibrium angle between the two even legs of the triangle $\alpha_e$, equilibrium leg length $R_{AB}$, well depth of the triatomic ion $D_e$, additive two-body part of the binding energy $D_{2b}$, nonadditive three-body part of the binding energy $D_{3b}$, and dissociation limit. \label{tab:ABC+_ALE_sym}} 
\begin{ruledtabular}
\begin{tabular}{llrrrrrr}
$A_2B^+$ & Geometry & $\alpha_e\,$(degrees) & $R_{AB}\,$(bohr) & $D_e\,$(cm$^{-1}$) & $D_{2b}\,$(cm$^{-1}$) & $D_{3b}\,$(cm$^{-1}$) & Diss.\\
\hline
Mg$_3^+$ & linear & 180.0 & 5.76 & 16408 & 21127 & -4720 & Mg$^+$+Mg+Mg \\
Mg$_2$Ca$^+$ & linear & 180.0 & 6.63 & 9200 & 10677 & -1477 & Ca$^+$+Mg+Mg \\
Mg$_2$Sr$^+$ & linear & 140.6 & 7.02 & 7418 & 8444 & -1026 & Sr$^+$+Mg+Mg \\
Mg$_2$Sr$^+$ & triangular & 65.2 & 6.85 & 7564 & 8781 & -1218 & Sr$^+$+Mg+Mg \\
Mg$_2$Ba$^+$ & linear & 169.3 & 7.41 & 6386 & 7130 & -744 & Ba$^+$+Mg+Mg \\
Mg$_2$Ba$^+$ & triangular & 58.8 & 7.00 & 7114 & 7237 & -123 & Ba$^+$+Mg+Mg \\
Mg$_2$Yb$^+$ & linear & 144.8 & 7.04 & 7169 & 8535 & -1367 & Yb$^+$+Mg+Mg \\
Ca$_2$Mg$^+$ & linear & 180.0 & 6.51 & 12201 & 6857 & 5344 & Ca$^+$+Ca+Mg \\
Ca$_3^+$ & linear & 180.0 & 7.34 & 14599 & 18611 & -4012 & Ca$^+$+Ca+Ca \\
Ca$_3^+$ & triangular & 79.0 & 7.14 & 13874 & 19326 & -5452 & Sr$^+$+Ca+Ca \\
Ca$_2$Sr$^+$ & linear & 180.0 & 7.72 & 11909 & 14544 & -2635 & Sr$^+$+Ca+Ca \\
Ca$_2$Sr$^+$ & triangular & 69.4 & 7.42 & 11826 & 15371 & -3546 & Sr$^+$+Ca+Ca \\
Ca$_2$Ba$^+$ & linear & 146.5 & 8.13 & 9933 & 11890 & -1957 & Ba$^+$+Ca+Ca \\
Ca$_2$Ba$^+$ & triangular & 62.1 & 7.67 & 11091 & 12546 & -1454 & Ba$^+$+Ca+Ca \\
Ca$_2$Yb$^+$ & linear & 180.0 & 7.69 & 11493 & 7899 & 3594  & Ca$^+$+Ca+Yb \\
Sr$_2$Mg$^+$ & linear & 180.0 & 6.86 & 10978 & 5898 & 5079 & Sr$^+$+Sr+Mg \\
Sr$_2$Ca$^+$ & linear & 180.0 & 7.68 & 12962 & 8729 & 4233 & Sr$^+$+Sr+Ca \\
Sr$_2$Ca$^+$ & triangular & 79.6 & 7.40 & 12218 & 14354 & -2136 & Sr$^+$+Sr+Ca \\
Sr$_3^+$ & linear & 167.9 & 8.08 & 13311 & 17160 & -3850 & Sr$^+$+Sr+Sr \\
Sr$_3^+$ & triangular & 74.4 & 7.74 & 13184 & 17982 & -4798 & Sr$^+$+Sr+Sr \\
Sr$_2$Ba$^+$ & linear & 148.1 & 8.47 & 11158 & 13643 & -2485 & Ba$^+$+Sr+Sr \\
Sr$_2$Ba$^+$ & triangular & 63.9 & 8.03 & 12208 & 14375 & -2167 & Ba$^+$+Sr+Sr \\
Sr$_2$Yb$^+$ & linear & 180.0 & 8.04 & 9919 & 6382 & 3536 & Sr$^+$+Sr+Yb \\
Ba$_2$Mg$^+$ & linear & 180.0 & 7.19 & 10737 & 5755 & 4982 & Ba$^+$+Ba+Mg \\
Ba$_2$Mg$^+$ & triangular & 95.7 & 6.84 & 9442 & 10051 & -609 & Ba$^+$+Ba+Mg \\
Ba$_2$Ca$^+$ & linear & 180.0 & 8.02 & 12085 & 7828 & 4257 & Ba$^+$+Ba+Ca \\
Ba$_2$Ca$^+$ & triangular & 78.8 & 7.59 & 11916 & 13798 & -1882 & Ba$^+$+Ba+Ca \\
Ba$_2$Sr$^+$ & linear & 180.0 & 8.43 & 12174 & 8473 & 3701 & Ba$^+$+Ba+Sr \\
Ba$_2$Sr$^+$ & triangular & 73.2 & 7.96 & 12395 & 14993 & -2599 & Ba$^+$+Ba+Sr \\
Ba$_3^+$ & linear & 158.3 & 8.85 & 13319 & 17195 & -3876 & Ba$^+$+Ba+Ba \\
Ba$_3^+$ & triangular & 65.5 & 8.26 & 14938 & 17911 & -2973 & Ba$^+$+Ba+Ba \\
Ba$_2$Yb$^+$ & linear & 180.0 & 8.42 & 8958 & 5574 & 3384 & Ba$^+$+Ba+Yb \\
Ba$_2$Yb$^+$ & triangular & 77.6 & 8.11 & 8279 & 11531 & -3251 & Ba$^+$+Ba+Yb \\
\end{tabular}
\end{ruledtabular}
\end{table*}

\begin{table*}[tbp!]
\caption{Characteristics of asymmetric forms of alkaline-earth-metal triatomic molecular ions $AAB^+$ in the lowest doublet $^2A_1$ electronic state which coreduces to the $^2\Sigma^+$ symmetry at a linear equilibrium geometry within the $C_{\infty v}$ point group. Consecutive columns list: equilibrium distance between the first and the second atom $R_{12}$, equilibrium distance between the second and the third atom $R_{23}$, well depth of the triatomic ion $D_e$, additive two-body part of the binding energy $D_{2b}$, nonadditive three-body part of the binding energy~$D_{3b}$, and dissociation limit. \label{tab:ABC+_ALE_asym}} 
\begin{ruledtabular}
\begin{tabular}{lrrrrrr}
$AAB^+$ & $R_{12}\,$(bohr) & $R_{23}\,$(bohr) & $D_e\,$(cm$^{-1}$) & $D_{2b}\,$(cm$^{-1}$) & $D_{3b}\,$(cm$^{-1}$) & Diss.\\
\hline
MgMgCa$^+$ & 5.98 & 6.45 & 8946 & 5436 & 3510 & Ca$^+$+Mg+Mg \\
MgMgSr$^+$ & 6.05 & 6.80 & 7253 & 4353 & 2900 & Sr$^+$+Mg+Mg \\
MgMgBa$^+$ & 6.13 & 7.13 & 6238 & 3744 & 2493 & Ba$^+$+Mg+Mg \\
MgMgYb$^+$ & 5.87 & 6.82 & 7904 & 4146 & 3758 & Yb$^+$+Mg+Mg \\
CaCaMg$^+$ & 7.24 & 6.78 & 12038 & 9765 & 2274 & Ca$^+$+Ca+Mg \\
CaCaSr$^+$ & 7.54 & 7.48 & 11819 & 8470 & 3349 & Sr$^+$+Ca+Ca \\
CaCaBa$^+$ & 7.49 & 7.94 & 9936 & 7054 & 2882 & Ba$^+$+Ca+Ca \\
CaCaYb$^+$ & 7.26 & 7.77 & 12901 & 9584 & 3317 & Ca$^+$+Ca+Yb \\
SrSrMg$^+$ & 7.94 & 7.23 & 10678 & 9115 & 1563 & Sr$^+$+Sr+Mg \\
SrSrCa$^+$ & 8.45 & 7.76 & 12518 & 9287 & 3231 & Sr$^+$+Sr+Ca  \\
SrSrBa$^+$ & 8.19 & 8.36 & 10956 & 7970 & 2986 & Ba$^+$+Sr+Sr  \\
SrSrYb$^+$ & 7.95 & 8.25 & 11257 & 8906 & 2351 & Sr$^+$+Sr+Yb \\
BaBaMg$^+$ & 8.74 & 7.62 & 10373 & 9238 & 1135 & Ba$^+$+Ba+Mg \\
BaBaCa$^+$ & 8.78 & 8.25 & 11811 & 9775 & 2036 & Ba$^+$+Ba+Ca \\
BaBaSr$^+$ & 8.80 & 8.55 & 12327 & 9789 & 2538 & Ba$^+$+Ba+Sr \\
BaBaYb$^+$ & 8.89 & 8.97 & 10676 & 9080 & 1595 & Ba$^+$+Ba+Yb \\
\end{tabular}
\end{ruledtabular}
\end{table*}
\begin{table*}[tbp!]
\caption{Characteristics of alkaline-earth-metal triatomic molecular ions $A_2B^+$ in the lowest doublet $^2B_2$ electronic state. All molecular ions have an isosceles triangular equilibrium geometry within the $C_{2v}$ point group. Consecutive columns list: equilibrium angle between the two even legs of the triangle $\alpha_e$, equilibrium leg length $R_{AB}$, well depth of the triatomic ion $D_e$, additive two-body part of the binding energy $D_{2b}$, nonadditive three-body part of the binding energy~$D_{3b}$, and dissociation limit. \label{tab:ABC+_ALE_B2}} 
\begin{ruledtabular}
\begin{tabular}{lrrrrrr}
$A_2B^+$ & $\alpha_e\,$(degrees) & $R_{AB}\,$(bohr) & $D_e\,$(cm$^{-1}$) & $D_{2b}\,$(cm$^{-1}$) & $D_{3b}\,$(cm$^{-1}$) & Diss. \\
\hline
Mg$_3$$^+$ & 42.3 & 7.47 & 13604 & 17861 & -4257 & Mg$^+$+Mg+Mg \\
Mg$_2$Ca$^+$ & 43.2 & 7.30 & 3964 & 7482 & -3518 & Ca$^+$+Mg+Mg \\
Mg$_2$Sr$^+$ & 40.3 & 7.70 & 1883 & 5177 & -3294 & Sr$^+$+Mg+Mg \\
Mg$_2$Yb$^+$ & 40.1 & 7.87 & 4314 & 5212 & -898 & Yb$^+$+Mg+Mg \\
Ca$_2$Mg$^+$ & 59.0 & 7.03 & 11939 & 14586 & -2647 & Ca$^+$+Ca+Mg \\
Ca$_3$$^+$ & 51.5 & 7.90 & 13662 & 18727 & -5065 & Ca$^+$+Ca+Ca \\
Ca$_2$Sr$^+$ & 49.6 & 8.16 & 11149 & 13510 & -2361 & Sr$^+$+Ca+Ca \\
Ca$_2$Ba$^+$ & 47.4 & 8.43 & 8075 & 10984 & -2909 & Ba$^+$+Ca+Ca \\
Ca$_2$Yb$^+$ & 49.3 & 8.34 & 12399 & 16329 & -3930 & Ca$^+$+Ca+Yb \\
Sr$_2$Mg$^+$ & 61.1 & 7.44 & 11176 & 12848 & -1672 & Sr$^+$+Sr+Mg \\
Sr$_2$Ca$^+$ & 58.9 & 7.55 & 12808 & 15905 & -3097 & Sr$^+$+Sr+Ca \\
Sr$_3$$^+$ & 54.4 & 8.32 & 13064 & 17673 & -4609 & Sr$^+$+Sr+Sr \\
Sr$_2$Ba$^+$ & 49.9 & 8.85 & 10087 & 12824 & -2737 & Ba$^+$+Sr+Sr \\
Sr$_2$Yb$^+$ & 54.0 & 8.51 & 11218 & 14282 & -3065 & Sr$^+$+Sr+Yb \\
Ba$_2$Mg$^+$ & 61.3 & 7.83 & 11995 & 11944 & 51 & Ba$^+$+Ba+Mg \\
Ba$_2$Ca$^+$ & 59.6 & 8.10 & 14015 & 14857 & -842 & Ba$^+$+Ba+Ca \\
Ba$_2$Sr$^+$ & 57.8 & 8.42 & 13891 & 15860 & -1969 & Ba$^+$+Ba+Sr \\
Ba$_3^+$ & 57.5 & 8.52 & 15382 & 17643 & -2261 & Ba$^+$+Ba+Ba \\
Ba$_2$Yb$^+$ & 58.7 & 8.45 & 12008 & 13163 & -1155 & Ba$^+$+Ba+Yb \\
\end{tabular}
\end{ruledtabular}
\end{table*}

We calculate the potential energy surfaces for all 25 alkali-metal triatomic molecular ions $A_2B^+$ consisting of Li, Na, K, Rb, Cs in the lowest singlet $^1A_1$ electronic state (ground state) at the isosceles triangular geometry and all 45 alkali-metal triatomic molecular ions in the lowest triplet $^3B_2$ electronic state which coreduces to the $^3\Sigma_u^+$ and $^3\Sigma^+$ symmetries at two possible linear geometries $ABA^+$ and $AAB^+$, respectively. The equilibrium geometries, that is the equilibrium angles between the two even legs of the triangle $\alpha_e$ and equilibrium leg lengths $R_{AB}$ for the isosceles triangular singlet states and the equilibrium distances between the first and the second atom $R_{12}$ and equilibrium distances between the second and the third atom $R_{23}$ for the linear triplet states, are collected in Table~\ref{tab:ABC+_AL_S} and Table~\ref{tab:ABC+_AL_T}, respectively. 

We also calculate the potential energy surfaces for 20 symmetric forms of alkaline-earth-metal triatomic molecular ions $ABA^+$ consisting of Mg, Ca, Sr, Ba, Yb in the lowest doublet $^2A_1$ electronic state at the isosceles triangular geometry and 16 asymmetric forms of alkaline-earth-metal triatomic molecular ions $AAB^+$ in the doublet $^2A_1$ electronic ground state which coreduces to the $^3\Sigma^+$ symmetry at the linear geometry. Additionally, we investigate the potential energy surfaces for 20 symmetric alkaline-earth-metal triatomic molecular ions $A_2B^+$ in the lowest doublet $^2B_2$ electronic state at the isosceles triangular geometry. Their equilibrium geometries are reported in Table~\ref{tab:ABC+_ALE_sym}, Table~\ref{tab:ABC+_ALE_asym}, and Table~\ref{tab:ABC+_ALE_B2}, respectively. Combinations containing more than one Yb atom are not presented because of numerical complexity and discrepancies between results obtained with different pseudopotentials. 

\begin{figure}[tb!]
\begin{center}
\includegraphics[width=\columnwidth]{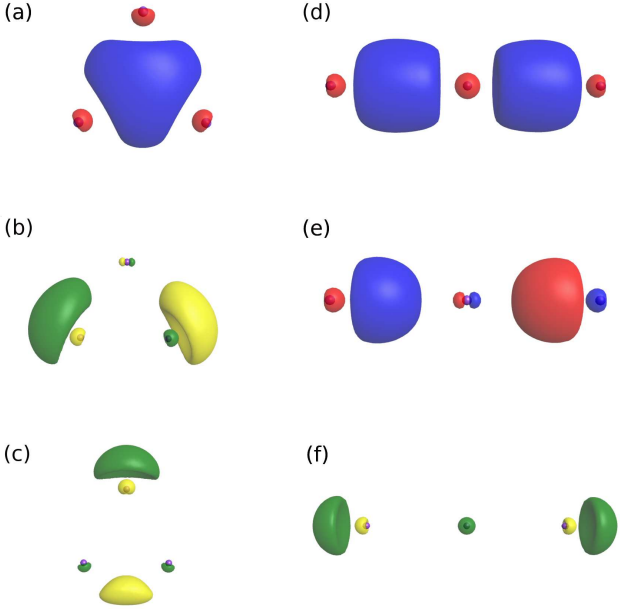}
\end{center}
\caption{Molecular orbital isosurfaces of homonuclear alkali-metal triatomic molecular ions in the singlet $^1A_1$ electronic state having the $^1A'_1$ symmetry within the $D_{3h}$ point group at the equilibrium triangular geometry: (a)~closed-shell $a_1'$ HOMO, (b)~$e'$ LUMO, (c)~$e'$ LUMO+1, and in the triplet $^3B_2$ electronic state having the $^3\Sigma^+_u$ symmetry within the $C_{\infty v}$ point group at the equilibrium linear geometry: (d)~open-shell $\sigma_g$ HOMO-1, (e)~open-shell $\sigma_u$ HOMO, (f)~$\sigma_g$ LUMO.}
\label{fig:MO_AL}
\end{figure}

\begin{figure}[tb!]
\begin{center}
\includegraphics[width=\columnwidth]{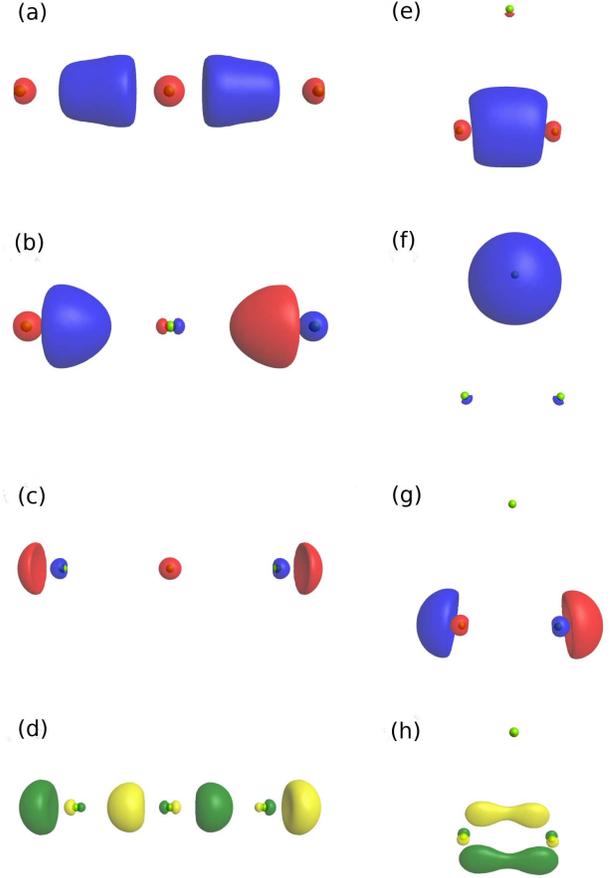}
\end{center}
\caption{Molecular orbital isosurfaces of homonuclear alkaline-earth-metal triatomic molecular ions in the doublet $^2A_1$ electronic state having the $^2\Sigma^+_g$ symmetry within the $C_{\infty v}$ point group at the equilibrium linear geometry: (a)~closed-shell $\sigma_g$ HOMO-2, (b)~closed-shell $\sigma_u$ HOMO-1, (c)~open-shell $\sigma_g$  HOMO, (d)~$\sigma_u$ LUMO, and in the doublet $^2B_2$ electronic state within the $C_{2v}$ point group at the equilibrium triangular geometry: (e)~closed-shell $a_1$ HOMO-2, (f)~closed-shell $a_1$ HOMO-1, (g)~open-shell $b_2$ HOMO, (h)~$b_1$ LUMO.
}
\label{fig:MO_AE}
\end{figure}

Together with equilibrium geometries, the well depths of the triatomic ions and their decomposition into additive two-body and nonadditive three-body parts are reported in Tables~\ref{tab:ABC+_AL_S}-\ref{tab:ABC+_ALE_B2}. The full potential well depth equals to the sum of all two-body and three-body contributions. Three-body nonadditive part, depending on its sign, may both stabilize and destabilize molecular ions. For molecular ions, and ion-neutral complexes in general, there is an ambiguity in the decomposition of the interaction energy into two-body or many-body contributions, because this decomposition depends on the formal assignment of the charge to one of the monomers. In the interacting ion-neutral systems, the charge can be delocalized or transferred to other monomer. The problem is minimized for systems where one of the monomers has significantly smaller ionization potential than others. In the present study, we assume that the charge is associated with the atom with the smallest ionization potential. For homonuclear molecular ions the most symmetric position of the charge is assumed. The interplay of the two-body and three-body interactions in the investigated molecular ions is discussed in the following paragraphs.

\begin{figure*}[tb!]
\begin{center}
\includegraphics[width=\columnwidth]{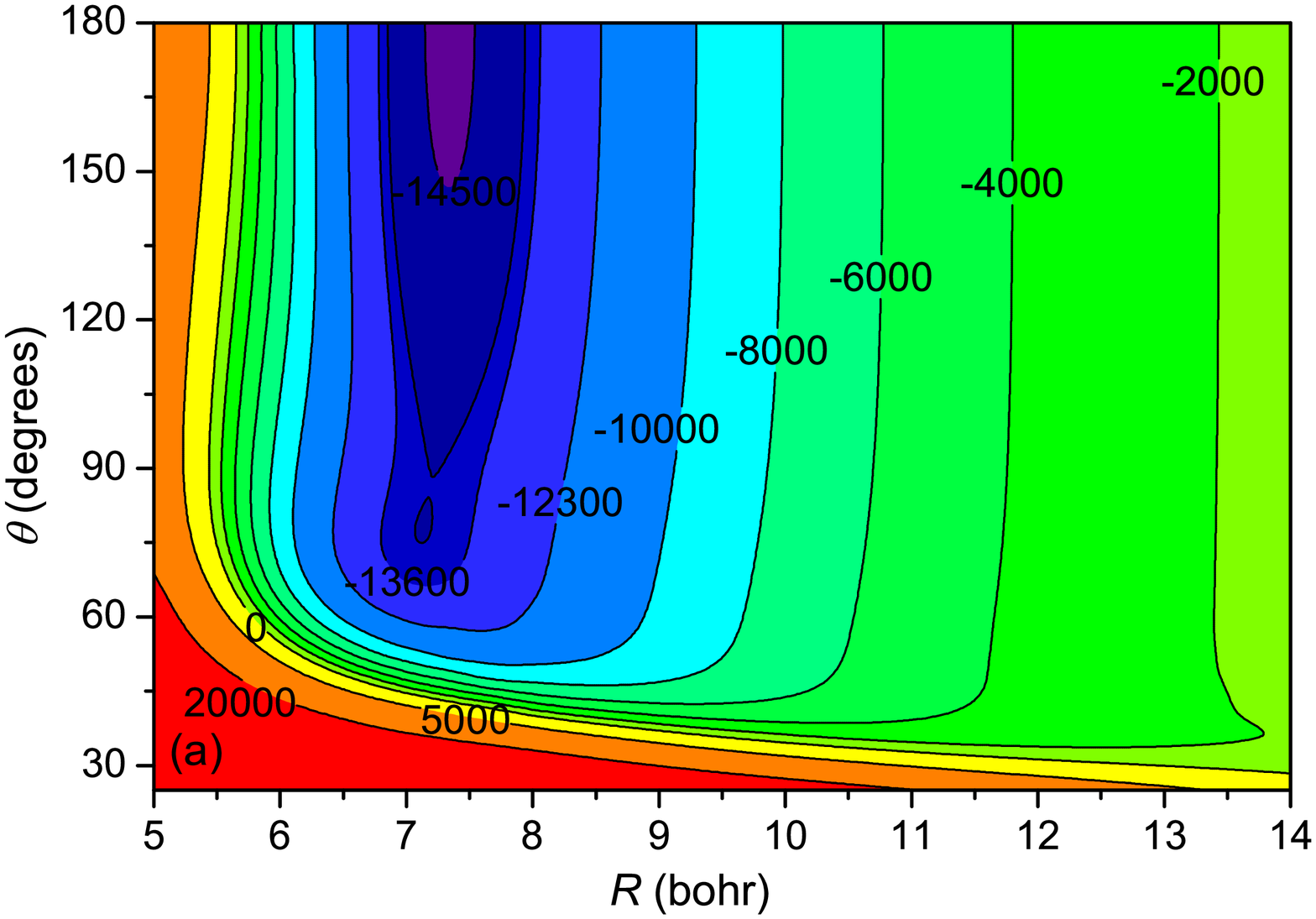}
\includegraphics[width=\columnwidth]{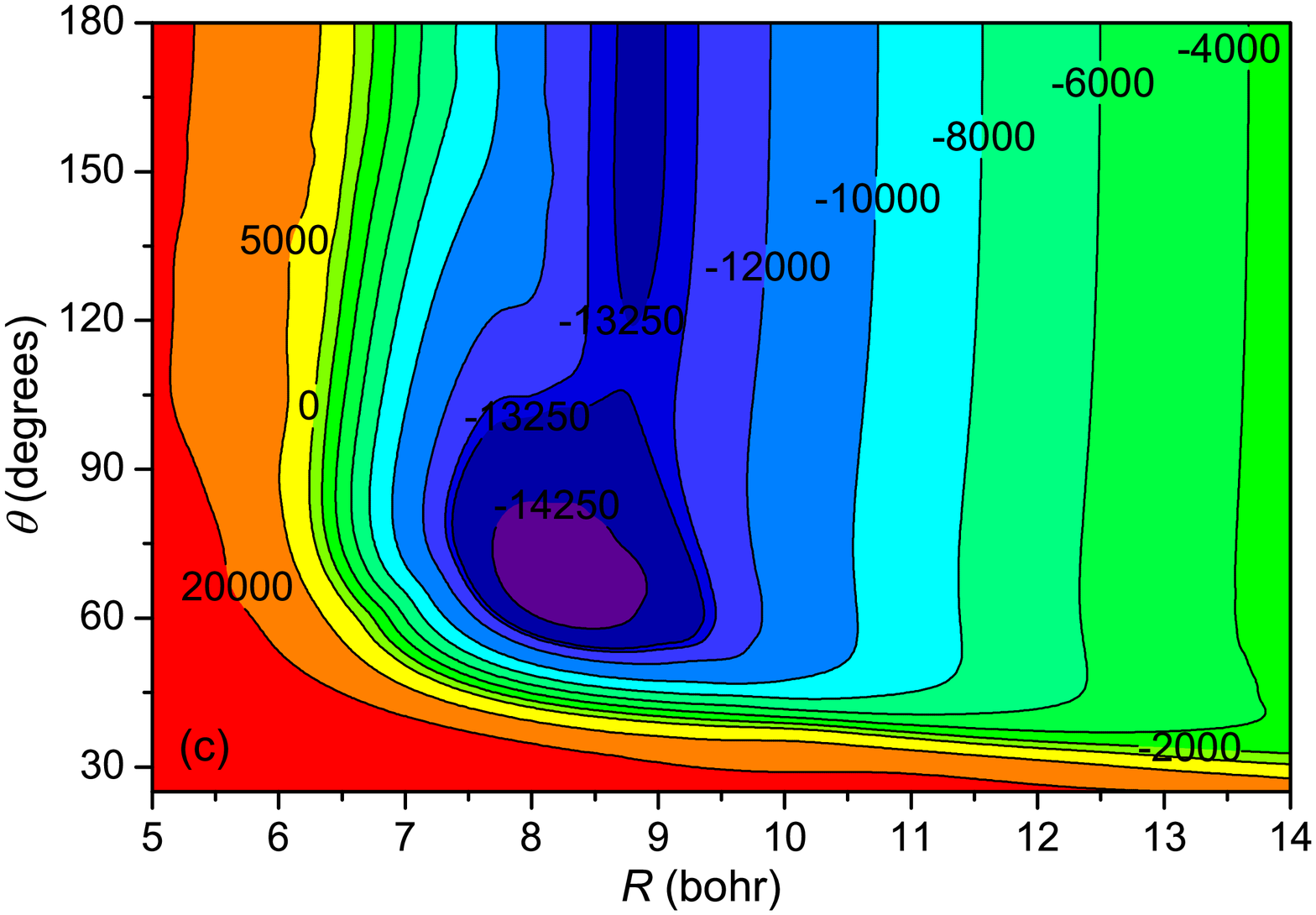}
\includegraphics[width=\columnwidth]{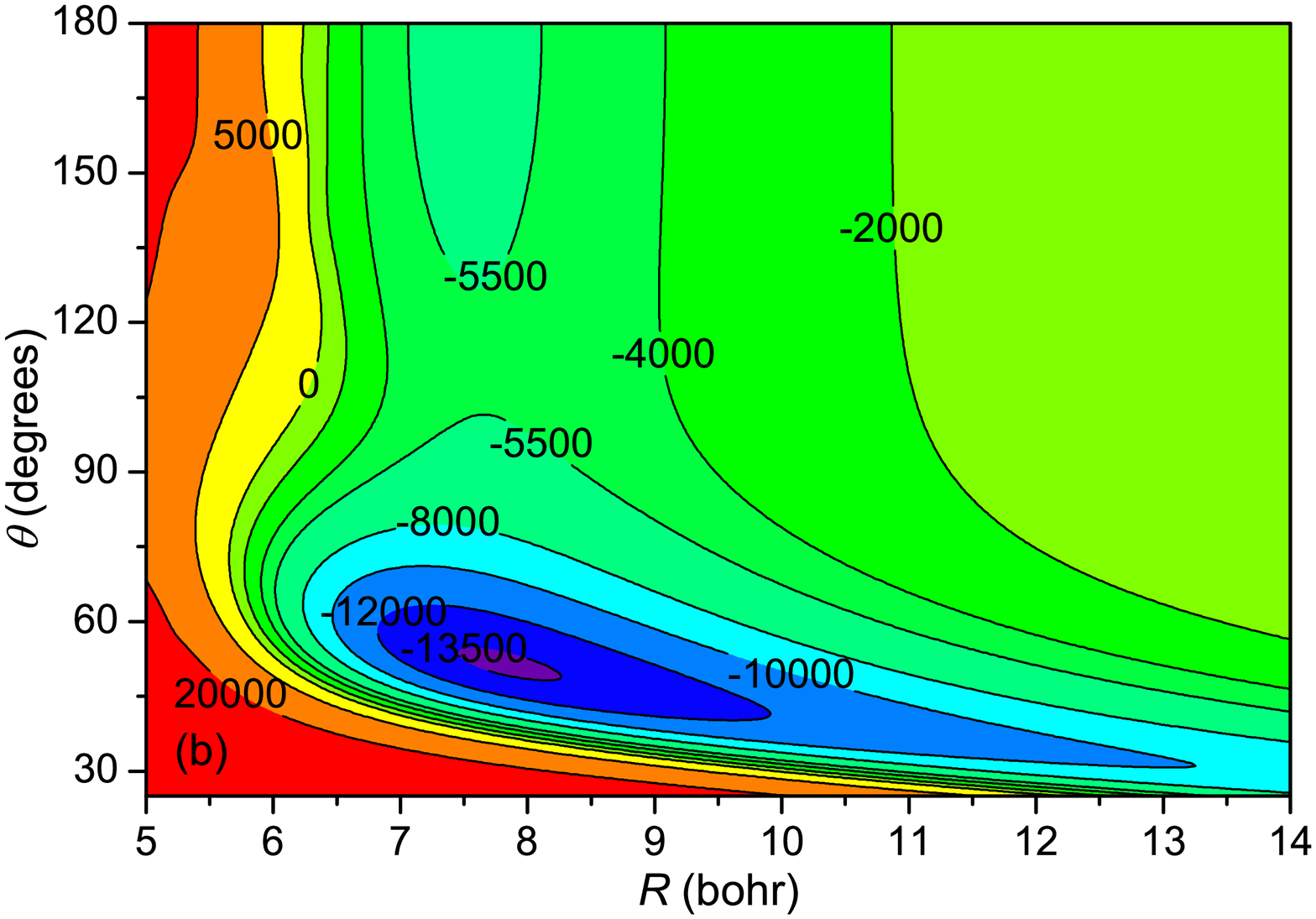}
\includegraphics[width=\columnwidth]{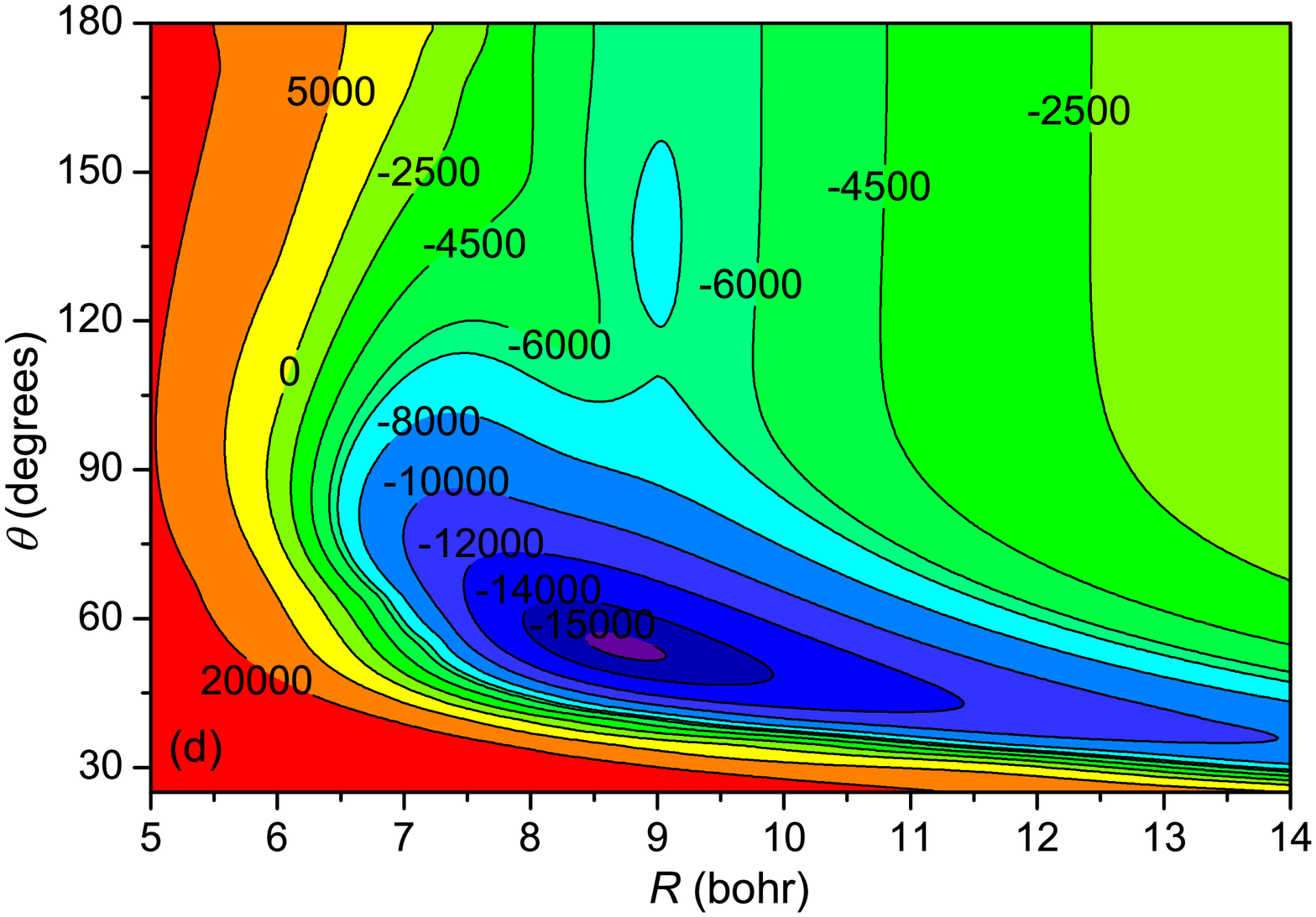}
\end{center}
\caption{Two-dimensional cuts through the ground-state potential energy surfaces of homonuclear alkaline-earth-metal triatomic molecular ions: (a)~Ca$_3^+$ in the $^2A_1$ electronic state, (b)~Ca$_3^+$ in the $^2B_2$ electronic state, (c)~Ba$_3^+$ in the $^2A_1$ electronic state, (d)~Ba$_3^+$ in the $^2B_2$ electronic state.}
\label{fig:2D}
\end{figure*}

All heteronuclear alkali-metal triatomic molecular ions $A_2B^+$ have the electronic ground state of the singlet $^1A_1$ symmetry at an isosceles triangular geometry with the $B$ on the symmetry axis within the $C_{2v}$ point group (see Table~\ref{tab:ABC+_AL_S}). Homonuclear ions $A_3^+$ have the electronic ground state of the singlet $^1A'_1$ symmetry at a equilateral triangular geometry within the $D_{3h}$ point group. The Na$_2$Li$^+$ molecular ion is the most strongly bound with $D_e=26004\,$cm$^{-1}$, followed by Li$_3^+$ and Li$_2$Na$^+$ with $D_e=23610$cm$^{-1}$ and $D_e=20281$cm$^{-1}$, respectively. 
The Na$_2$Cs$^+$, K$_2$Cs$^+$, and Rb$_2$Cs$^+$ molecular ions are the most weakly bound with $D_e=11240$cm$^{-1}$, $D_e=11495$cm$^{-1}$, and $D_e=11502$cm$^{-1}$, respectively. Interestingly, the three-body interaction destabilizes all alkali-metal triatomic molecular ions except for the most strongly bound Na$_2$Li$^+$, which is stabilized by $D_{3b}=$3470$\,$cm$^{-1}$. On the other hand, the Li$_3^+$ molecular ion with the largest stabilizing two-body term has also the largest destabilizing three-body term of $D_{3b}=$-5190$\,$cm$^{-1}$. The present predictions of equilibrium geometries agree with previous works on alkali triatomic molecular ions~\cite{PavoliniJCP87,SpiegelmannJCP88,KhannaPRL88,JeungCPL90,HeerRMP93,WangTCA94,SmartJMS96,LyalinPRA07}, while previously predicted well depths are usually underestimated probably because they were obtained using smaller basis sets and lower level methods.

Equilibrium internuclear distances between any two of the three atoms in singlet-state alkali-metal triatomic molecular ions are generally smaller than in the corresponding ionic and neutral dimers. The shortening is typically around 0.1-0.3$\,$bohr, with the average of 0.25$\,$bohr for homonuclear ions, and the largest values of 0.68$\,$bohr for Cs$_2$Na$^+$ and Cs$_2$Li$^+$. The largest shortening is observed for systems with the smallest destabilizing effect of the three-body interaction. Equilibrium angles $\measuredangle ABA^+$ range from 38 degrees for Li$_2$Cs$^+$ to 101 degrees for Cs$_2$Li$^+$ and correlate with the size of involved atoms, i.e.~they are acute when $m_A<m_B$ and obtuse when $m_A>m_B$.

All heteronuclear alkali-metal triatomic molecular ions in the lowest triplet electronic state have two linear equilibrium geometries: symmetric $ABA^+$ and asymmetric $AAB^+$ ones, separated by an energy barrier of few thousands cm$^{-1}$ (see Table~\ref{tab:ABC+_AL_T}). The symmetry of the electronic wave function for the symmetric equilibrium is $^3\Sigma^+_u$ while it is $^3\Sigma^+$ for asymmetric one. The LiLiLi$+$ molecular ion is the most strongly bound with $D_e=16856\,$cm$^{-1}$, while the NaCsNa$^+$ molecular ion is the most weakly bound with $D_e=5512$cm$^{-1}$. If the ionization potential of $A$ is larger than that of $B$, then the asymmetric $AAB^+$ equilibrium geometry has a smaller energy, otherwise the symmetric $ABA^+$ geometry is a global minimum of the triplet potential energy surface. Interestingly, the three-body interaction destabilizes all triplet-state homonuclear alkali-metal triatomic molecular ions (e.g.~Li$_3^+$ by $D_{3b}=-4158\,$cm$^{-1}$), whereas its effect for heteronuclear $ABA^+$ and $AAB^+$ ions depends on the ionization potentials of $A$ and $B$ atoms. If the ionization potential of $A$ is smaller than of $B$, then the three-body interaction stabilizes both $ABA^+$ and $AAB^+$ ions, otherwise the three-body interaction stabilizes asymetric $AAB^+$ ions but destabilizes symmetric $ABA^+$ ones. The largest stabilizing three-body energy term of $D_{3b}=4842\,$cm$^{-1}$ is for the asymmetric LiLiNa$^+$ molecular ion, while the largest destabilizing three-body energy term of $D_{3b}=-6179\,$cm$^{-1}$ is for the symmetric NaRbNa$^+$ molecular ion. Equilibrium internuclear distances between any two of the three atoms in triplet-state alkali-metal triatomic molecular ions are generally larger than in the corresponding ionic dimers.

To explain the different triangular and linear equilibrium geometries of the lowest singlet and triplet electronic states of the alkali-metal triatomic molecular ions, in Fig.~\ref{fig:MO_AL}, we plot exemplary isosurfaces of highest occupied molecular orbitals (HOMOs) and lowest unoccupied molecular orbitals (LUMOs) for a homonuclear case of these ions. Molecular orbitals for heteronuclear ions look very similar with small alterations due to the broken symmetry. In the singlet state, two valence electrons occupy single orbital which is highly bonding due to a significant charge delocalization and large electron density between nuclei (see Fig.~\ref{fig:MO_AL}(a)). Two plotted LUMOs are antibonding and degenerate for homonuclear ions. In the triplet state, however, one electron has to be excited from the lowest valence orbital. The second valence orbital has antibonding character at triangular geometry (see Fig.~\ref{fig:MO_AL}(b)), whereas it starts to be bonding at the linear geometry (see Fig.~\ref{fig:MO_AL}(e)). Thus, at the linear geometry, two occupied orbitals in the triplet state have bonding character with large electron densities between nuclei that stabilize the  triplet-state molecular ions.

Alkaline-earth-metal triatomic molecular ions $A_2B^+$ have five valence electrons occupying two closed-shell and one open-shell valence orbitals in the lowest doublet electronic states, therefore these ions have a richer structure of possible equilibrium geometries as compared with alkali-metal ions. Their electronic ground state can have either doublet $^2A_1$ or $^2B_2$ symmetry within the $C_{2v}$ point group symmetry. Interestingly, for both electronic symmetries there may exist two minima, the first one at an isosceles triangular geometry with the $B$ atom on the symmetry axis within the $C_{2v}$ point group symmetry, and the second one at a linear or close to linear geometry.

Among homonuclear triatomic alkaline-earth-metal molecular ions $A_3^+$, Mg$_3^+$, Ca$_3^+$, and Sr$_3^+$ have their ground state of the $^2A_1$ symmetry, while Ba$_3^+$ has the $^2B_2$ symmetry ground state. In the $^2A_1$ state, Mg$_3^+$ has a single global minimum of the $^2\Sigma_g^+$ symmetry at the linear geometry, while Ca$_3^+$, Sr$_3^+$, Ba$_3^+$ have two local minima. For Ca$_3^+$, the minimum at the linear geometry in the global minimum. For Sr$_3^+$ and Ba$_3^+$ there are two local minima at triangular geometries, one with acute and one with obtuse (close to 180 degrees) angles. In the $^2A_1$ state, the global minimum of Sr$_3^+$ is at an obtuse angle while the global minimum of Ba$_3^+$ is at an acute angle. PESs for Ca$_3^+$ and Ba$_3^+$ in the $^2A_1$ and $^2B_2$ electronic states are presented in Fig.~\ref{fig:2D}. For both ions and both states, two local minima are clearly visible. In the $^2A_1$ state, two minima have similar well depths, while in the $^2B_2$ state, the minima at the acute triangular geometry have much larger well depths. Therefore in Table~\ref{tab:ABC+_ALE_B2} we report global minima of the $^2B_2$ state only. 

Heteronuclear alkaline-earth-metal molecular ions $A_2B^+$ have an additional local minimum at the asymmetric $AAB^+$ linear geometry of the $^2A_1$ state which coreduces to the $^2\Sigma^+$ symmetry and has a similar well depth as other minima. The Sr$_2$Mg$^+$, Sr$_2$Yb$^+$, Ba$_2$Mg$^+$, Ba$_2$Ca$^+$, Ba$_2$Sr$^+$, and Ba$_2$Yb$^+$ molecular ions have their ground state of the $^2B_2$ symmetry at the acute triangular geometry. All remaining heteronuclear alkaline-earth-metal trimers, notably all of the compounds with two Mg or Ca atoms, have their ground state of the $^2A_1$ symmetry with a variety of geometries from symmetric linear to acute or obtuse triangular ones and asymmetric linear ones. Interestingly, all heteronuclear molecular ions containing one Yb atom have their ground state at the asymmetric linear geometry. Among all alkaline-earth-metal triatomic molecular ions, the Mg$_3^+$ ion in the $^2A_1$ state, is the most strongly bound with $D_e=16408\,$cm$^{-1}$, followed by other homonuclear ions in the $^2A_1$ and $^2B_2$ states. On the other hand, heteronuclear alkaline-earth-metal molecular ions containing two Mg atoms are the most weakly bound. The Mg$_2$Ba$^+$ ion at the symmetric close-to-linear geometry in the $^2A_1$ state has the well depth of $D_e=6386\,$cm$^{-1}$, the same MgMgBa$^+$ ion at the asymmetric linear geometry  in the $^2A_1$ state has a well depth of $D_e=3744\,$cm$^{-1}$ and Mg$_2$Ba$^+$ does not have a minimum in the $^2B_2$ state.

Equilibrium internuclear distances in alkaline-earth-metal triatomic molecular ions are smaller or larger, than equilibrium distances of the corresponding ionic and neutral dimers, depending on triatomic ions's electronic state and geometry. For the linear geometry, both symmetric and asymmetric molecular ions generally have larger equilibrium distances. For the isosceles triangular geometry, the molecular ions in the $^2A_1$ electronic state have the distance between A and B atoms (the legs of the triangle) smaller while between two A atoms (the base of the triangle) larger as compared with the corresponding ionic dimers, while the molecular ions in the $^2B_2$ electronic state shown the opposite pattern.      

To understand the different equilibrium geometries of the alkaline-earth-metal triatomic molecular ions, in Fig.~\ref{fig:MO_AE}, we plot exemplary isosurfaces of valence orbitals for a homonuclear case of these ions in the $^2A_1$ electronic state at the linear geometry and in the $^2B_2$ electronic state at the triangular geometry. Molecular orbitals for heteronuclear ions look similar with small alterations due to the broken symmetry. Interestingly, the molecular orbitals of the $^2A_1$ electronic state having the $^2\Sigma_g^+$ symmetry at the linear geometry, resemble the molecular orbitals of the alkali-metal triatomic molecular ions in the triplet $^3B_2$ electronic state having the $^2\Sigma_u^+$ symmetry at the same geometry. Two lowest closed-shell valence orbitals have bonding character with large electron densities between nuclei that stabilize the linear molecular ions (see Fig.~\ref{fig:MO_AE}(a) and Fig.~\ref{fig:MO_AE}(b)). The bonding character of the second valence orbital is reduced for triangular geometries, thus the linear minimum of the $^2A_1$ electronic state is relatively stabilized. On the other hand, the electron delocalization is reduced in molecular valence orbitals for alkaline-earth-metal triatomic molecular ions in the $^2B_2$ electronic state. In fact, the structure of molecular orbitals for this state can be well described as a sum of molecular orbitals of an $A_2^+$ diatomic molecular ion in the $^2\Sigma_u$ electronic state and a ground-state $B$ atom. Thus no proper chemical bonding is present in the $^2B_2$ electronic state and the acute triangular geometry is favored for this state because of steric effects. The interplay of the structure of molecular orbitals and the correlation of valence electrons results in two local minima for most of symmetric alkaline-earth-metal triatomic molecular ions in contrast to alkali-metal ions.

Interestingly, homonuclear alkaline-earth-metal triatomic molecular ions have a permanent electric dipole moments along their symmetry axis. In the $^2B_2$ electronic state, Mg$_3^+$, Ca$_3^+$, Sr$_3^+$, and Ba$_3^+$ have permanent electric dipole moments of 1.71$\,$D, 0.33$\,D$, 0.51$\,$D, and 0.77$\,$D, respectively. Obtained permanent electric dipole moments are in agreement with the picture presented using molecular orbital description of $^2B_2$-state molecular ions as $A_2^++B$ in the previous paragraph. Additionally, calculations of partial electric charges, e.g.~using Hirshfeld analysis, confirm the favored charge localization at $A_2^+$. Heteronuclear alkaline-earth-metal triatomic molecular ions have larger permanent electric dipole moments of the order of magnitude as in diatomic molecular ions.

Three-body interaction destabilizes all alkaline-earth-metal triatomic molecular ions in the $^2A_1$ electronic state at the acute triangular geometry and almost all ions in the $^2B_2$ electronic state with the exception of Ba$_2$Mg$^+$. The largest destabilizing three-body energy terms are for homonuclear ions with the largest value of $D_{3b}=-5452\,$cm$^{-1}$ for Ca$_3^+$ in the $^2A_1$ state at the triangular geometry. Interestingly, for the $^2A_1$ electronic state at the linear geometry, the effect of three-body interactions is analogous as for alkali-metal molecular ions in the $^2A_1$ at the same geometry. That is the three-body interaction stabilizes all the asymmetric $AAB^+$ ions while the symmetric $ABA^+$ ones are stabilized only if the ionization potential of $A$ is larger than that of $B$. 

Unfortunately, for alkaline-earth-metal triatomic molecular ions, the ambiguity in the decomposition of the interaction energy into two-body or three-body contributions is larger than for alkali-metal ions. In fact, the sign of three-body energies can change for some ions, if other assignment of the charge to the monomers is employed. The most striking examples are homonuclear triangular ions in the $^2A_1$ electronic state which do not form equilateral triangles and for which the three-body energy changes by few thousands cm$^{-1}$ for different charge assignments. Detailed studies of unambiguous many-body energy decomposition in ionic systems are out of the scope of the present work.

\subsection{Chemical reactions in ion-neutral systems}

\begin{figure}[tb!]
\begin{center}
\includegraphics[width=\columnwidth]{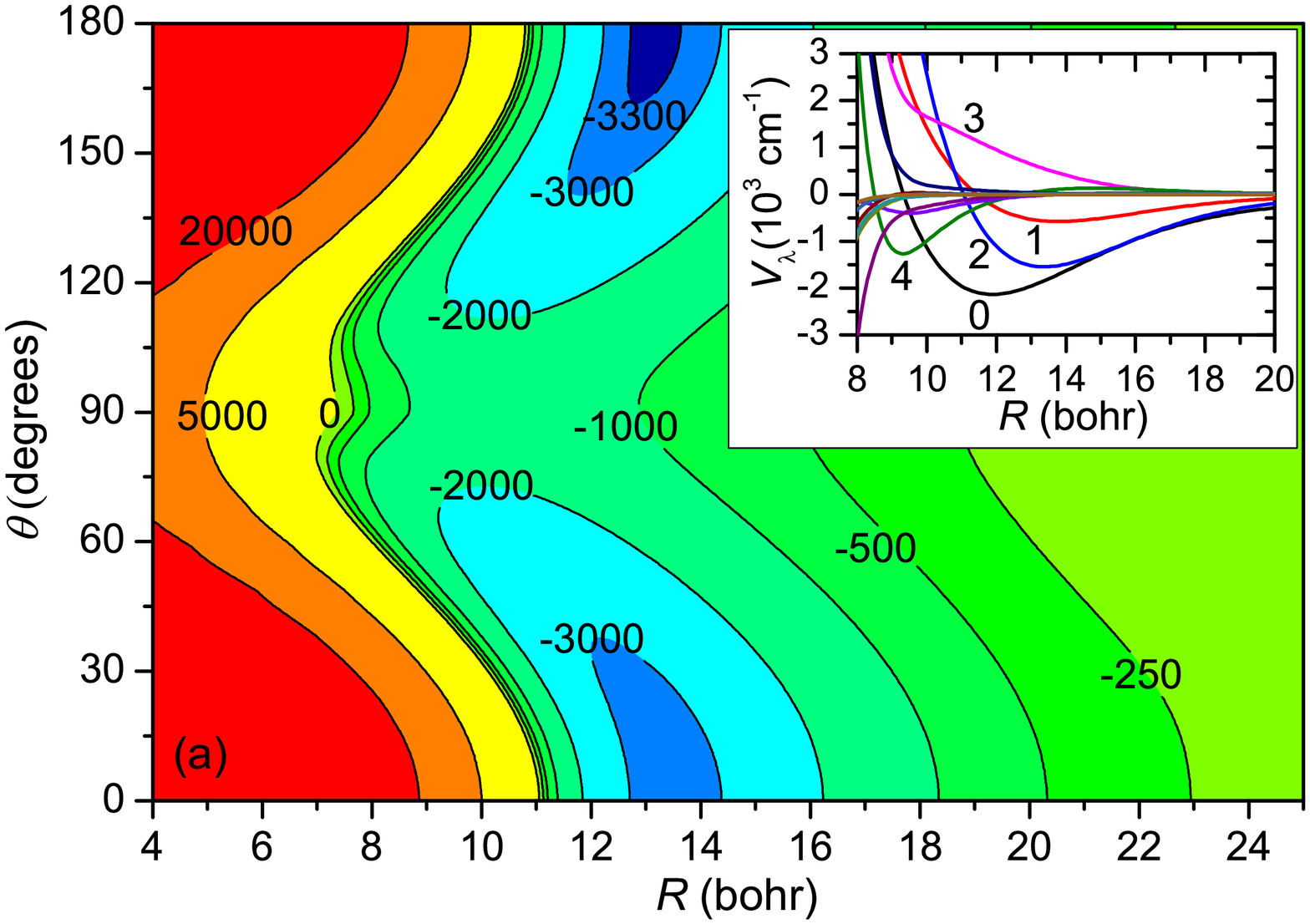}
\includegraphics[width=\columnwidth]{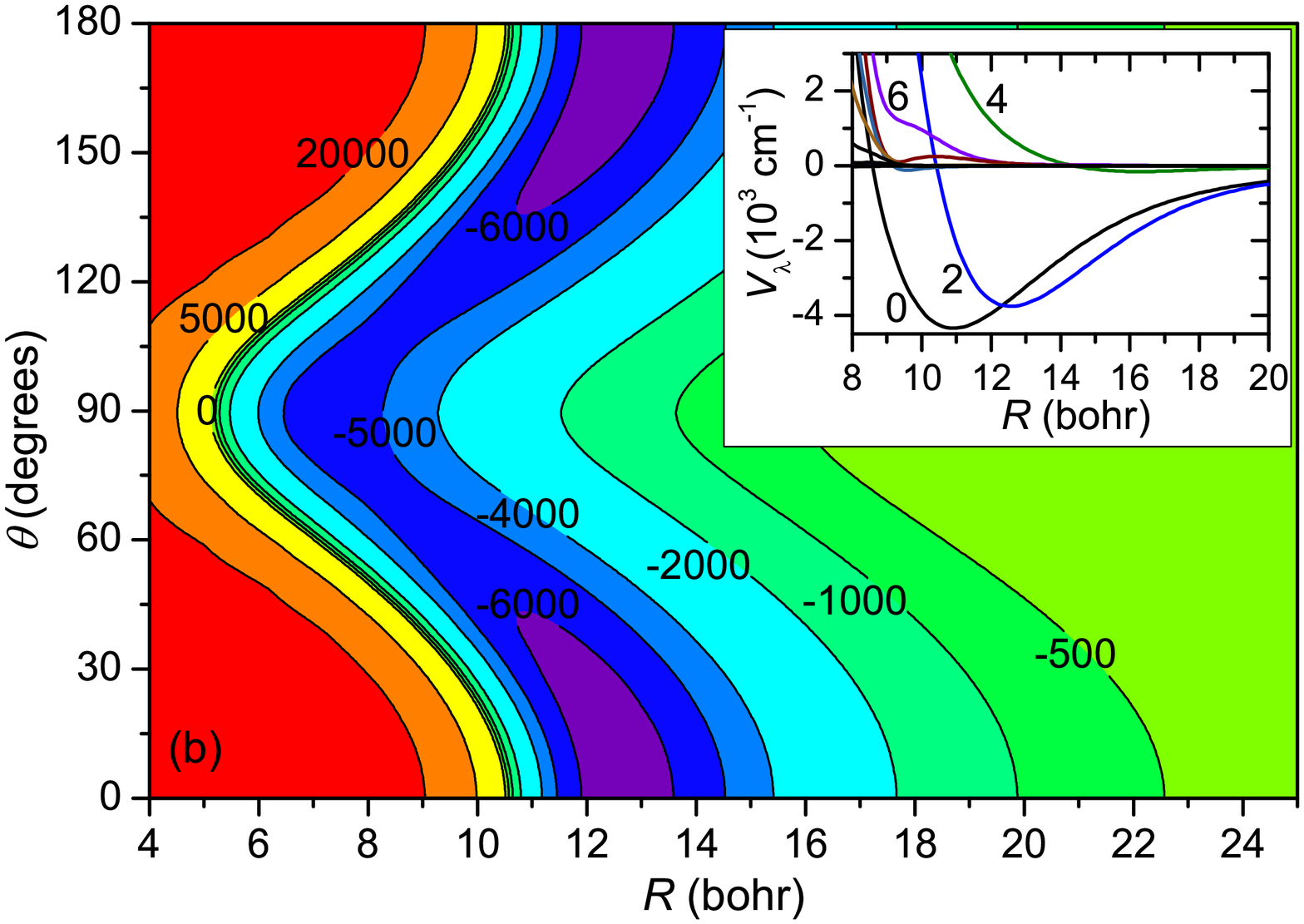}
\end{center}
\caption{Two-dimensional cuts through the ground-state potential energy surfaces of (a)~KRb$^+$+K and (b)~Rb$^+$+Sr$_2$ ion-neutral systems. Insets show the corresponding Legendre components.}
\label{fig:mol+at}
\end{figure}

The prospects for sympathetic cooling and applications of molecular ions immersed into ultracold atomic gases or atomic ions immersed into ultracold molecular gases can be jeopardized by possible chemical reactions, on one hand. However cold and controlled chemical reactions in these systems can be an interesting subject of study on its own, on the other hand.~\cite{DeiglmayrPRA12,TomzaPRL15,PuriScience17,KilajNC18,TomzaPCCP17,PuriNatChem19,Dorfler2019,KasPRA19}.

To study chemical reactions, potential energy surfaces of adequate quality are needed. In Fig.~\ref{fig:mol+at}, we present exemplary non-reactive two-dimensional PESs within the rigid rotor approximation for high-spin interaction between the KRb$^+$ molecular ion and K atom, and between the Rb$^+$ atomic ion and Sr$_2$ molecule. PESs are showed together with their decompositions onto Legendre components. Both surfaces are very anisotropic, with the first and second anisotropic Legendre component almost as large as the isotropic one. This suggests large inelastic rate coefficients for collisional rotational relaxation in the considered systems. PESs for other non-reactive systems using the presented methodology can be computed upon request. We will study the potential energy surfaces for reactive ion-neutral collisions in triatomic alkali-metal and alkaline-earth-metal systems in the future, while in the next paragraphs we analyze possible channels of chemical reactions based on the energetics of the reactants.

In the considered diatomic ion-atom systems there can be several paths of collision- and interaction-induced chemical reactions related to the charge rearrangement~\cite{TomzaRMP19}. The radiative charge transfer (RCT)
\begin{equation}\label{eq:RCT}
A^+ + B \to A + B^+ + \hbar\omega\,,
\end{equation}
where the electron is spontaneously transferred from the $B$ atom to the $A^+$ ion emitting a photon of energy $\hbar\omega$. This process is possible when the ionization potential of the neutral $A$ atom is not smaller than the ionization potential of the $B$ atom~\cite{TomzaPRA15b}, 
\begin{equation}
IP(A)\geq IP(B)\,.
\end{equation}
The non-radiative charge transfer (nRCT) driven by non-adiabatic couplings can also be possible for the same energetic condition if electronic states associated with $A^++B$ and $A+B^+$ thresholds form an avoided crossing at shorter internuclear distances. The radiative association (RA)
\begin{equation}\label{eq:RA}
A^+ + B \to AB^+(v,j) + \hbar\omega\,,
\end{equation}
where the $A^+$ ion and the $B$ atom spontaneously form a $AB^+$ molecular ion in a $(v,j)$ ro-vibrational state emitting a photon of energy~$\hbar\omega$. Such a process driven by the transition between two electronic states is possible when the reaction \eqref{eq:RCT} is energetically allowed or when the interaction energy in the $AB^+$ molecular ion is larger or equal to the missing difference of the ionization potentials. The spontaneous radiative association is also possible (but very unlikely) for all polar complexes $AB^+$ driven by the transition between ro-vibrational levels of the electronic ground state. Finally, the spontaneous radiative deexcitation of the formed $AB^+$ molecular ion in an excited $(j,v)$ ro-vibrational level to lower-energy levels is feasible
\begin{equation}\label{eq:deex}
AB^+(v,j) \to AB^+(v',j\pm 1) + \hbar\omega\,.
\end{equation}
Energetics of reactions given by Eq.~\eqref{eq:RCT} and Eq.~\eqref{eq:RA} can be assigned by using the data from Table~\ref{tab:atoms}.

In the considered triatomic ion-atom systems, combinations of ultracold molecular ions with atoms, $AB^++C$, or atomic ions with molecules, $A^++BC$, lead to a broad range of possible chemical reactions related to the charge or atom rearrangement. Their possibility and energetics can be assigned by using the atomic and molecular data provided in the present work and collected in Tables~\ref{tab:atoms}-\ref{tab:ABC+_ALE_B2} together with previous results for neutral molecules~\cite{ZuchowskiPRA10,TomzaPRA13b}.

The non-radiative (or radiative) charge transfer
\begin{equation}\label{eq:RCT_mol}
\begin{split}
AB^+ + C &\to AB + C^+\ (+ \hbar\omega)\,, \\
A^+ + BC &\to A + BC^+\ (+ \hbar\omega)\,,
\end{split}
\end{equation}
is possible when  the electron attachment energy of the $AB^+$ molecular ion or $A^+$ ion is not smaller than the ionization potential of the neutral $C$ atom or $BC$ molecule,
\begin{equation}
\begin{split}
EA(AB^+) &\geq IP(C)\,, \\
EA(A^+) &\geq IP(BC)\,,
\end{split}
\end{equation}
where $EA(A^+)=IP(A)$, while $EA(AB^+)\approx IP(AB)$ for systems with diagonal Franck-Condon factors between levels of $AB^+$ and $AB$.
The radiative association is also possible for the same energetic conditions,
\begin{equation}\label{eq:RA_mol}
\begin{split}
AB^+ + C &\to ABC^+ + \hbar\omega\,, \\
A^+ + BC &\to ABC^+ + \hbar\omega\,.
\end{split}
\end{equation}

The non-radiative (or radiative) ion-exchange reaction
\begin{equation}\label{eq:ion-ex}
AB^+ + C \to A+BC^+\ (+ \hbar\omega)
\end{equation}
is possible when the dissociation energy of $AB^+$ is not larger than the dissociation energy of $BC^+$, provided lack of charge transfer between $B^+$ and $C$,
\begin{equation}
D_0(BC^+)\geq D_0(AB^+) \quad \text{for} \ IP(B)\leq IP(C)\,.
\end{equation}
If the charge transfer between $B^+$ and $C$ is possible, the above condition has to be corrected
\begin{equation}
\begin{split}
D_0(BC^+)+IP(B)-IP(C) & \geq D_0(AB^+) \\ 
& \text{for} \ IP(B) > IP(C)\,.
\end{split}
\end{equation}

The radiative or non-radiative  atom-exchange reaction
\begin{equation}\label{eq:at-ex}
AB^+ + C \to A^++BC\ (+ \hbar\omega)
\end{equation}
is possible when the dissociation energy of $AB^+$ is not larger than the dissociation energy of $BC$, provided the ionization potential of $A$ is not larger than the ionization potential of $B$, 
\begin{equation}\label{eq:at-ex_c1}
D_0(BC)\geq D_0(AB^+) \quad \text{for} \ IP(A)\leq IP(B)\,.
\end{equation}
If the ionization potential of $A$ larger than of $B$, the above condition has to be corrected
\begin{equation}\label{eq:at-ex_c2}
\begin{split}
D_0(BC^+) \geq   D_0(AB^+) & + IP(A)-IP(B) \\ 
& \text{for} \ IP(A) > IP(B)\,.
\end{split}
\end{equation}
The reverse reaction to the one given by Eq.~\eqref{eq:at-ex} is possible for reversed conditions of Eqs.~\eqref{eq:at-ex_c1}-\eqref{eq:at-ex_c2}.

\begin{figure}[tb!]
\begin{center}
\includegraphics[width=\columnwidth]{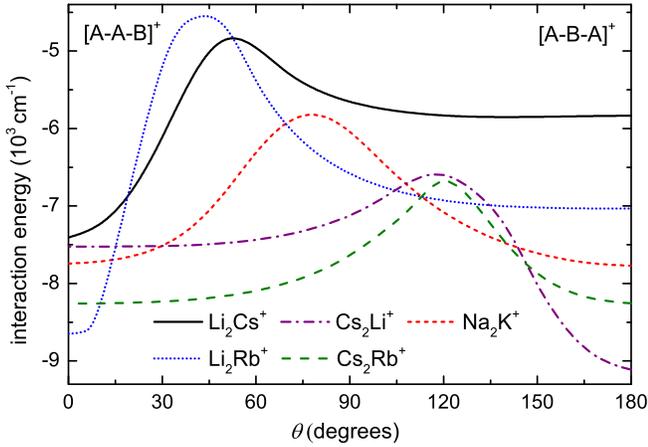}
\end{center}
\caption{Minimum-energy reaction paths for the isomerisation of selected alkali-metal triatomic molecular ions between their asymmetric $AAB^+$ and symmetric $ABA^+$ forms.}
\label{fig:izo}
\end{figure}

The collisional dissociation associated with the charge transfer
\begin{equation}\label{eq:dis}
\begin{split}
AB^+ + C \to A+B+C^+\,, \\
A^+ + BC \to A+B+C^+
\end{split}
\end{equation}
is possible when the ionization potential of $A$ is larger than the ionization potential of $C$ by more than the dissociation energy of $AB^+$ or $BC$
\begin{equation}\label{eq:dis_c}
\begin{split}
IP(A)-IP(C)\geq D_0(AB^+)\,, \\
IP(A)-IP(C)\geq D_0(BC) \,.
\end{split}
\end{equation}

In the case of ro-vibrationally exited molecular ions, the collision-induced deexcitation may also happen 
\begin{equation}\label{eq:coldeex}
AB^+(v,j) + C \to AB^+(v',j') + C\,.
\end{equation}

In the case of linear molecular ions, the collision-induced isomerisation is possible 
\begin{equation}\label{eq:iso}
ABA^+(\{v\},j) + C \leftrightarrow AAB^+(\{v'\},j') + C\,,
\end{equation}
where the reaction direction depends on which isomer has a larger energy. The minimum-energy reaction paths for the spontaneous isomerisation of selected alkali-metal triatomic molecular ions are presented in Fig.~\ref{fig:izo}. The spontaneous isomerisation is strongly suppressed because of the large energy barriers, nevertheless the heights of those barriers give a good estimation of the interaction energy needed to allow for collision-induced isomerisation.

The intermolecular charge-transfer reactions
\begin{equation}
AB + CD^+ \to  AB^+ + CD\,,
\end{equation}
are possible when the ionization potential of $AB$ is not smaller than the ionization potential of $CD$. The energetics of the atom- and ion-exchange reactions 
\begin{equation}
\begin{split}
AB + CD^+ &\to AC + BD^+\,,\\
AB + CD^+ &\to AC^+ + BD\,,
\end{split}
\end{equation}
can also be assigned using the dissociation and ionization energies of involved molecules and molecular ions.

If molecular reactants are not in the ground vibrational level ($v$=0), then their dissociation energies $D_0$ in all the above conditions should be replaced by dissociation energies for considered vibrational level, $D_v$. If the reaction energy is smaller than the uncertainty of the calculated dissociation and ionization energies, then predicted above energetics may be less accurate.

If the above considered reactions are energetically forbidden, they may potentially be induced by the laser-field excitation of involved reactants~\cite{SullivanPRL12,Dorfler2019,PuriNatChem19}. If the excitation energy to the lowest excited electronic state of one of atomic or molecular reactants is larger than the missing reaction energy then the endoenergetic chemical reaction on the ground potential energy surface becomes exoenergetic on the electronically-excited potential energy surface~\cite{PuriScience17}.

Finally, the presented atomic and molecular data also allows to assign energetics of ionic chemical reactions
\begin{equation} 
\begin{split}
AB^+ + AB^+ & \to A_2^+ + B_2^+ \,, \\
AB^+ + CD^+ & \to AC^+ + BD^+ \,, \\
AB^+ + C^+ & \to A^+ + BC^+\,, \\ 
\end{split}
\end{equation}
however such reactions are not relevant for ultracold systems due to highly repulsive nature of the Coulomb interaction.

\section{Summary and Conclusions}
\label{sec:summary}

Experiments employing ultracold molecular ions and ion-neutral mixtures are a promising platform to further our understanding of physical basis of chemistry and to perform high precision measurements essential for testing fundamental laws of nature.  Compared to neutral molecules, molecular ions are easier to prepare, trap, and detect. In this paper, we have presented theoretical results for all ground-state diatomic $AB^+$ and most of triatomic $A_2B^+$ molecular ions  consisting of alkali-metal and alkaline-earth-metal atoms. We have employed \textit{ab initio} techniques of quantum chemistry, such as the coupled cluster method restricted  to  single,  double,  and  noniterative  triple  excitations,  CCSD(T), combined with large Gaussian basis sets and small-core energy-consistent pseudopotentials, to obtain equilibrium distances, atomization energies, ionization potentials, permanent electric dipole moments, and polarizabilities.

We have predicted a wide range of dissociation energies and permanent electric dipole moments for the dimers, and a variety of equilibrium geometries for the trimers from equilateral triangular through isosceles triangular to linear. We have also evaluated and characterized three-body nonaddtive interactions in these systems at equilibrium geometries. We have identified possible channels of chemical reactions in ionic two-body: $A^+$+B and $AB^+$, and three-body systems: $A^++AB$, $AB^++A$, and $A_2B^+$, based on the energetics of the reactants. Additional, we have provided two-dimensional interaction potential energy surfaces for KRb$^+$+K and Sr$_2$+Rb$^+$ mixtures and we have presented example calculations of minimum energy paths for the isomerisation reaction of linear alkaline-metal trimers in the lowest triplet electronic state between asymmetric $AAB^+$ and symmetric $ABA^+$ arrangements. PESs for other systems can be computed upon request.

The present results may be useful for investigating controlled chemical reactions and other applications of alkali and alkaline-earth molecular ions immersed in or formed from ultracold gases. Collected molecular properties can be employed in quantum scattering calculations or easily used to access chemical characteristics of selected systems realized in modern cold hybrid ion-neutral experiments. In the future, we plan to use the calculated potential energy surfaces to study two-body and three-body collisions in ion-neutral systems and to extend presented results for three-body interactions of atomic and molecular ions with atoms or molecules in excited electronic states. 

\begin{acknowledgments}
We would like to thank Tatiana Korona for many useful discussions and help with the \textsc{Molpro} and Gaussian programs. We acknowledge support from the National Science Centre Poland (2016/23/B/ST4/03231 and 2015/19/D/ST4/02173) and the PL-Grid Infrastructure.
\end{acknowledgments}

\bibliography{ABC}

\end{document}